\documentclass[12pt]{article}
\usepackage{amsmath}
\usepackage{amssymb}
\begin{document}
\begin{center}
{\bf {\large{Massive Spinning  Relativistic Particle: Revisited Under\\
BRST and Supervariable  Approaches}}}

\vskip 2.5cm

{\sf  A. Tripathi$^{(a)}$, B. Chauhan $^{(a)}$, A. K. Rao $^{(a)}$,  R. P. Malik$^{(a,b)}$}\\
$^{(a)}$ {\it Physics Department, Institute of Science,}\\
{\it Banaras Hindu University, Varanasi - 221 005, (U.P.), India}\\

\vskip 0.1cm

$^{(b)}$ {\it DST Centre for Interdisciplinary Mathematical Sciences,}\\
{\it Institute of Science, Banaras Hindu University, Varanasi - 221 005, India}\\
{\small {\sf {e-mails: ankur1793@gmail.com; bchauhan501@gmail.com;\\
amit.akrao@gmail.com;  rpmalik1995@gmail.com}}}
\end{center}

\vskip 1.5 cm

\noindent
{\bf Abstract:}
We discuss the continuous and infinitesimal gauge, supergauge, reparameterization,  nilpotent Becchi-Rouet-Stora-Tyutin (BRST)
and anti-BRST symmetries and derive corresponding nilpotent charges for the one (0+1)-dimensional (1D) {\it massive} model of a
spinning relativistic particle. We exploit the theoretical potential and power of the BRST and supervariable approaches
to derive the (anti-)BRST symmetries and coupled (but equivalent) Lagrangians for this system. In particular, we capture
the off-shell nilpotency and absolute anticommutatvity of the  conserved (anti-)BRST charges within the framework of the {\it newly}
proposed (anti-)chiral supervariable approach (ACSA) to BRST formalism where {\it only} the (anti-)chiral supervariables
(and their suitable  super expansions are taken into account along the Grassmannian direction(s)). One of the novel observations of our 
present investigation is the derivation of the Curci-Ferrari (CF)-type restriction by the requirement of the absolute
anticommutatvity of the (anti-)BRST charges in the {\it ordinary} space. We obtain the {\it same} restriction within the 
framework of ACSA to BRST formalism by (i) the symmetry invariance of the coupled  Lagrangians, and (ii) the 
proof of the absolute anticommutatvity of the conserved and nilpotent (anti-)BRST charges. The observation of
the anticommutativity property of the (anti-)BRST charges is a novel result in view of the fact that we have 
taken into account {\it only} the (anti-)chiral super expansions.



\newpage

\section {Introduction}

\noindent
The basic concepts behind the {\it local} gauge theories are at the heart of a precise theoretical description of {\it three} out of
{\it four} fundamental interactions of nature. Becchi-Rouet-Stora-Tyutin (BRST) formalism [1-4] is one of the most intuitive and beautiful
approaches to quantize the {\it local} gauge theories where the unitarity and {\it quantum} gauge [i.e. (anti-)BRST]
invariance are respected {\it together} at any arbitrary order of perturbative computations for a given physical process 
that is permitted by the local (i.e. interacting) gauge theory at the {\it quantum} level. A couple of decisive 
features of the BRST formalism are the nilpotency of the (anti-)BRST symmetries as well as the existence of the  absolute
anticommutativity property between the BRST and anti-BRST symmetry transformations for a given local {\it classical} gauge  
transformation. The hallmark of the {\it quantum} (anti-)BRST symmetries is the existence of the (anti-)BRST invariant 
Curci-Ferrari (CF)-type restriction(s) [5, 6] that ensure the absolute anticommutativity property of the (anti-)BRST 
symmetry transformations {\it and} the existence of the coupled (but equivalent) Lagrangian densities for the {\it quantum}
gauge theories. The Abelian 1-form gauge theory is an exception where the CF-type restriction is {\it trivial} and the 
Lagrangian density is {\it unique} (but {\it that} is a limiting case of the non-Abelian 1-form gauge theory where 
the CF-condition [7] exists).

The usual superfield approach (USFA) to BRST formalism [8-15] sheds light on the {\it geometrical} origin for the 
off-shell nilpotency and absolute anticommutativity of the (anti-) BRST symmetry transformations where the
horizontality condition (HC) plays an important and decisive role [10-12]. These approaches, however, lead to the 
derivation of the (anti-)BRST symmetries for the gauge field and associated (anti-)ghost fields {\it only} [10-12].
The above USFA does {\it not} shed any light on the (anti-)BRST symmetries, associated with the {\it matter} fields,
in an {\it interacting} gauge theory. In our earlier works (see, e.g. [16-19]), we have systematically and consistently {\it 
generalized} the USFA where, in addition to the HC, we exploit the potential and power of the gauge invariant restrictions
(GIRs) to obtain the (anti-)BRST symmetry transformations for the {\it matter}, (anti-)ghost and gauge fields of an 
{\it interacting} gauge theory {\it together}. There is {\it no} conflict between the HC and GIRs as they compliment
and supplement each-other in a beautiful fashion. This approach has been christened  as the augmented version of
superfield approach\footnote{The USFA with HC, developed in [10-12], is mathematically very elegant and, in one stroke,
it leads to the derivation of proper (anti-)BRST symmetry transformations for the gauge and
associated (anti-)ghost fields along with the derivation of the (anti-)BRST invariant CF-condition. The AVSA is
a minor extension of [10-12] where HC and gauge invariant restrictions are exploited  {\it together} (cf. Sec. 8 below).} (AVSA) to BRST formalism.

In the recent set of papers [20-24], we have developed a {\it simpler} version of the AVSA where {\it only} the
(anti-)chiral supervariables/superfields and {\it their} appropriate super expansions have been taken into consideration.
This superfield approach to BRST formalism has been christened  as the (anti-)chiral superfield/supervariable approach (ACSA).
It may be mentioned here that, in {\it all} the {\it earlier} superfield approaches [8-19], the {\it full} super expansions
of the superfields/supervariables, along {\it all} the Grassmannian directions of the (D, 2)-dimensional supermanifold, 
have been taken into account for the consideration of a D-dimensional {\it local} gauge invariant theory (defined on the {\it flat}
Minkowskian  space). One of the decisive features of the ACSA to BRST formalism is {\it its} dependence on the {\it quantum}
gauge [i.e. (anti-)BRST] invariant restrictions on the supervariables/superfields which lead to the {\it derivation}
of appropriate (anti-)BRST symmetry transformations for {\it all} the fields/variables of the theory {\it together}
with the deduction of the (anti-)BRST invariant CF-type restriction(s). The upshot of the results from ACSA is the 
observation that the conserved and nilpotent (anti-)BRST charges turn out to be absolutely anticommuating in nature
{\it despite} the fact that {\it only} the (anti-)chiral super expansions of the supervariables/superfields are
taken into account (within the framework of ACSA to BRST formalism).

The purpose of our present investigation is to apply the ACSA to BRST formalism to the 1D system of a {\it massive} spinning relativistic particle
and derive the proper (anti-) BRST symmetry transformations for {\it this} system so that it can be discussed and described within the framework
of BRST formalism. Our present 1D reparameterization invariant system is important in its own right as it provides a prototype model for the
(super)gauge invariant theory as well as an example for a toy model of supergravity theory. Needless to say, its generalization leads to
the theory of superstrings, too. If the existence of the continuous symmetries is the {\it guiding} principle for the definition of a {\it beautiful}
theory in physics, the 1D model of a {\it massive} spinning particle {\it represents} one such example which encompasses in its folds a host of 
beautiful continuous symmetries (cf. Secs. 2, 3 below). In our present investigation, we lay a whole lot of emphasis on the off-shell nilpotent 
and absolutely anticommuting (anti-)BRST symmetry transformations of our 1D system and derive the corresponding conserved Noether charges.
It is worthwhile to mention, at this stage, that {\it physically} the property of off-shell nilpotency of the (anti-)BRST symmetries
and corresponding charges imply their {\it fermionic} nature and the absolute anticommutativity property  encodes the linear 
{\it independence} of the above nilpotent symmetries and charges.

Against the backdrop of the above discussions,
in our present endeavor, we have shown the existence of the {\it three classical} level symmetries which are the gauge, supergauge and
reparameterization transformations [cf. Eqs. (2), (4)] under which the first-order Lagrangian $(L_f)$ for the 1D system of a {\it massive} spinning
relativistic particle remains 
invariant. We have further established that the reparameterization symmetry transformations contain (i) the 
{\it gauge} symmetry transformations [cf. Eq. (2)] provided {\it all} the fermionic variables are set equal to zero
(i.e. $\psi_\mu = \psi_5 = \chi = 0$), and (ii) the combination of gauge and supergauge symmetry transformations
[cf. Eq. (7)] {\it under} specific {\it conditions} where the appropriate equations of motion and identifications of the
transformation parameters are taken into account [cf. Eqs. (6), (9), (10)]. We have elevated the 
{\it classical} (super)gauge symmetry transformations (7) to the {\it quantum} level within the framework 
of BRST formalism and derived the (anti-) BRST symmetries that are respected  by the coupled (but equivalent)
Lagrangians $L_b$ and $L_{\bar b}$ [cf. Eqs. (17), (18)]. We have  demonstrated that {\it both} the
Lagrangians are {\it equivalent} because {\it both} of them respect {\it both} the BRST and anti-BRST 
symmetry transformations at the {\it quantum} level provided the whole theory is considered  on the
sub-manifold of the {\it quantum} Hilbert space of variables where the CF-type restriction
is satisfied [cf. Eqs. (20)-(23)]. We have further shown the existence of the (anti-)BRST invariant CF-type  restriction at 
the level of the proof of absolute anticommutativity of  the (anti-)BRST conserved charges in the {\it ordinary} space [cf. Eqs. (35)-(38)].

In our present endeavor, we have captured {\it all} the above {\it key features} within the framework
of ACSA to BRST formalism where {\it only} the (anti-)chiral supervariables and their corresponding super expansion(s)
along the Grassmannian direction(s) of the (1, 1)-dimensional (anti-)chiral super sub-manifolds of the 
{\it general} (1, 2)-dimensional supermanifold have been taken into consideration in a consistent and systematic
fashion. One of the novel observations is the proof of the absolute anticommutativity property 
of the conserved and nilpotent (anti-)BRST charges within the ambit  of ACSA to BRST formalism where {\it only}
the (anti-)chiral super expansion(s) of the (anti-) chiral supervariables have been taken into account. Moreover, 
we note that the {\it above} proof {\it also} distinguishes between the {\it chiral} and {\it anti-chiral}
(1, 1)-dimensional super sub-manifolds {\it within} the framework of ACSA to BRST formalism (cf. Appendix D below).

Our present investigation is essential and interesting on the following counts. First and foremost, our 1D system of the 
{\it massive} spinning relativistic particle is more {\it general} than its {\it massless} counterpart
which has been discussed in our earlier work [25]. Second, our present system is a toy model of
a supersymmetric gauge theory whose generalization to 4D provides a model for the supergravity
theory  with a cosmological constant term. Hence, this toy  model is interesting and important in  its own right. 
Third, our present model is {\it also} a generalization of the  scalar relativistic particle where the 
{\it fermionic} as well as {\it bosonic} (anti-)ghost variables appear within the framework of BRST formalism.
Fourth, we have been curious to find out the contribution  of the mass-term 
(and its associated variable) in the determination of the gauge-fixing and
Faddeev-Popov ghost terms within the framework  of BRST formalism [cf. Eq. (16) below]. Fifth, we have found
out the CF-type restriction for the 1D {\it massless} spinning particle in our earlier work by exploiting 
the beauty of the super-symmetrization of horizontalitiy condition [25]. Thus, we are now curious to find out {\it its}
existence by proving the absolute anticommutativity of the conserved (anti-)BRST charges. 
Furthermore, we are interested  in capturing {\it its} existence within the framework of ACSA 
to BRST formalism. We have accomplished {\it all} these goals in our present endeavor. Finally, a {\it thorough} study of our 1D 
system of  a {\it massive} spinning relativistic has been a challenge for us as we have {\it already} studied  a scalar
relativistic  particle and a {\it massless} spinning relativistic particle from various angels in our earlier
works [25-32].

The theoretical material of our present endeavor is organized as follows. In Sec. 2, we discuss the gauge,
supergauge and reparameterization symmetries of the Lagrangian that describes the 1D {\it massive} spinning
relativistic particle. Our Sec. 3 deals with the (anti-) BRST symmetries corresponding to the {\it combined}
gauge and supergauge symmetries where the {\it fermionic} as well as the {\it bosonic} (anti-)ghost variables appear 
in the BRST analysis. The subject matter of Sec. 4 concerns itself with the derivation of the BRST symmetries
within the framework of ACSA to BRST formalism where the {\it quantum gauge} (i.e. BRST) invariant restrictions
 on the {\it anti-chiral} supervariables play a crucial role. Our Sec. 5 is devoted  to the derivation of
anti-BRST symmetries by exploiting the anti-BRST invariant restrictions  on the {\it chiral} supervariables
within the purview of ACSA to BRST formalism. In Sec. 6, we prove the existence of the CF-type restriction by 
capturing the {\it symmetry invariance} of the Lagrangians within the ambit of ACSA. We capture the off-shell 
nilpotency  and absolute anticommutatvity of the conserved (anti-)BRST charges by applying the  key techniques 
of ACSA to BRST formalism in Sec. 7. Finally, in Sec. 8, we make some concluding remarks and point 
out a few future directions for further investigations.

In our Appendices  A, B and C, we collect a few of the explicit computations which supplement as well as  complement 
some of the  crucial and key statements that have been made and emphasized in the main body of our present endeavor.
Our Appendix D is devoted to the discussion of an alternative proof of the absolute anticommutativity  of the 
(anti-)BRST charges {\it and} the existence of the CF-type restriction (i) in the {\it ordinary} space, and (ii) in the {\it superspace}
by exploiting the theoretical tricks and techniques of  ACSA.\\

\noindent
{\it Convention and Notations}: The free (i.e. ${\dot p}_\mu = 0$) massive spinning relativistic particle 
is embedded in a D-dimensional flat Minkowskian spacetime manifold  that is characterized by a metric tensor $\eta_{\mu\nu}$ 
= diag$\,(+1, -1, -1 ...)$ where the Greek indices $\mu, \nu, \lambda .... = 0, 1, 2...(D - 1)$.
We adopt the convention of the left-derivative w.r.t. the fermionic variables $(\chi, \psi_\mu, \psi_5, 
\gamma, c, {\bar c})$. We denote the (anti-)BRST fermionic $(s_{(a)b}^2 = 0)$ symmetry
transformations by the symbol $s_{(a)b}$ which {\it anticommutes} (i.e. $\chi \, s_{(a)b}
 = - \, s_{(a)b} \, \chi, \; s_{(a)b} \, \psi_\mu = - \, \psi_\mu \, s_{(a)b},$ etc.) 
 with {\it all} the fermionic variables $(\chi, \psi_\mu, \psi_5, \gamma, c, {\bar c})$ and
 {\it commutes} (i.e. $x_\mu \,s_{(a)b} = s_{(a)b}\,x_\mu, \;p_\mu \, s_{(a)b} = s_{(a)b}\,p_\mu, 
 \; e \,s_{(a)b} = s_{(a)b} \,e, \;  s_{(a)b} \,\bar b  = \bar b \, s_{(a)b},$ etc.)
 with {\it all} the bosonic variables  of our theory.  We  also denote the (anti-)BRST charges by the symbol  $Q_{(a)b}$.\\

\section{Preliminaries: Some Continuous and Infinitesimal  Symmetries in Lagrangian Formulation}

In this section, we discuss some {\it infinitesimal} and {\it continuous} symmetries and demonstrate their {\it equivalence} under
some specific conditions where the usefulness of some appropriate equations of motion as well as identifications of a few
transformations parameters has been exploited. We begin with the following {\it three} equivalent Lagrangians which describe the
1D system of a {\it massive} spinning relativistic particle (see, e.g. [33])
\begin{eqnarray}
&&L_0 = m\,\sqrt {(\dot x^\mu\,+i\,\chi \,\psi^\mu)(\dot x_\mu\,+i\,\chi\, \psi_\mu)} - i\,m\,\chi\,\psi_5,\nonumber\\ 
&&L_f = p_\mu\,\dot x^\mu +\frac{i}{2}(\psi_\mu \,\dot\psi^\mu - \psi_5 \,\dot \psi_5) - \frac {e}{2}\;(p^2 - m^2)
+ i\,\chi \,(p_\mu \,\psi^\mu - m\,\psi_5), \nonumber\\ 
&&L_s  = \frac {1}{2\,e}\;(\dot x_\mu + i\,\chi \,\psi_\mu)(\dot x^\mu + i\,\chi\, \psi^\mu) 
+ \frac {e}{2}\;m^2 +  \frac{i}{2}\,(\psi_\mu \,\dot\psi^\mu - \psi_5\, \dot \psi_5) - i\, m\,\chi\, \psi_5,
\end{eqnarray}
where $L_0$ is the Lagrangian with a square-root, $L_f$ is the first-order Lagrangian\footnote{We would
like to point out that, in Ref. [33], the emphasis is laid on the first-order Lagrangians and
their usefulness. Hence, the first-order Lagrangian ($L_f$) is the {\it only} Lagrangian that is mentioned in [33].} and $L_s$ is the 
second-order Lagrangian. Our one (0+1)-dimensional (1D) system is embedded in a flat Minkowskian D-dimensional
 target space  where ($ x_\mu,\,p^\mu $) are the  canonically conjugate {\it bosonic} 
coordinates and momenta (with $\mu = 0, 1, 2...D-1$). The trajectory of the particle is parameterized by an evolution parameter $ \tau$ and
generalized velocities ($ \dot x_\mu \,=\,\frac{d x_ \mu} {d\tau}\,, \dot\psi_\mu\,=\,\frac{d\psi_\mu} {d\tau}$)
are defined w.r.t. {\it it}. We have  fermionic $( \psi_\mu\, \psi_\nu + \psi_\nu \,\psi_\mu = 0, \;\chi\, \psi_\mu 
+ \psi_\mu \,\chi = 0,\; \psi_\mu \,\psi_5 + \psi_5 \,\psi_\mu = 0,\, \chi \,\psi_5 + \psi_5 \,\chi = 0, 
\chi^2 = 0, etc.)$ variables in our theory which commute $(\psi_\mu\, e - e \,\psi_\mu = 0,\psi_\mu \,x_\nu 
- x_\nu \,\psi_\mu = 0,\,\psi_\mu \,p _\nu - p _\nu \,\psi_\mu = 0,\;\dot x_\nu\, \psi_\mu - \psi_\mu\, \dot x_\nu = 0, \, etc.)$ 
with {\it all} the bosonic variables $(x_\mu, p_\mu, e )$ of our theory. It should be noted that $ \psi_\mu$  is the 
superpartner of $x_\mu$ and $\psi_5 $ variable has been invoked in the theory to incorporate a {\it mass} term 
$m$ so that the mass-shell condition $(p^2 - m^2 = 0)$ for the free particle could  be satisfied.

The Lagrangian $L_0$ has a square-root and its {\it massless} limit is {\it not} defined. On the other hand,
the second-order Lagrangian $(L_s)$ is endowed with a variable (i.e. einbein) which is located in the denominator.
Thus, the Lagrangians $ L_0 $ and $ L_s $ have their own limitations. We shall focus on the first-order Lagrangian
$(L_f)$ for our discussions where variables $ e (\tau)$ and $ \chi (\tau)$ are {\it not } purely Lagrange  multiplier variables 
but their  transformations are such that they behave like the ``gauge" and ``supergauge" variables [cf. Eq. (2) below].
Our 1D system is a model of supersymmetric gauge theory and its generalization to 4D theory provides a model for the 
supergravity theory where $\psi_\mu$ corresponds to the Rarita-Schwinger field and $e(\tau)$ becomes the vierbein field. The mass $m$,
in the supergravity theory, represents the cosmological constant term. In a nut-shell, our present 1D model of a {\it massive}
spinning relativistic particle is important and interesting  in its own right because its generalization {\it also}
becomes a model of the superstring theory (see, e.g. [34, 35]).

The Lagrangian $L_f$  respects the following gauge $(\delta_g)$ and supergauge $(\delta_{sg})$ symmetry transformations, namely;
\begin{eqnarray}
&&\delta_g x_\mu  = \xi\, p_\mu, \;\;\; \delta_g p_\mu = 0,\;\;\; \delta_g e  = \dot \xi,\;\;\; \delta_g \psi_\mu 
= 0,\;\;\; \delta_g \psi_5 = 0,\;\;\;\delta_g \chi = 0,\nonumber\\
&&\delta_{sg} \, x_\mu = \kappa \, \psi_\mu, \qquad \qquad \delta_{sg} \, \psi_\mu = i\,\kappa \, p_\mu, \qquad \qquad \delta_{sg} \, p_\mu = 0,\nonumber \\ 
&&\delta_{sg} \,\psi_5 = i\,\kappa \,m, \qquad \qquad  \delta_{sg} \, \chi = i\,\dot\kappa, \qquad \qquad \delta_{sg} \, e = 2\,\kappa \, \chi,
\end{eqnarray}
where $\xi(\tau)$ and $\kappa(\tau)$ are the {\it infinitesimal} gauge and supergauge symmetry
transformation parameters, respectively. It is straightforward to note that $\xi(\tau)$ is a
bosonic and $\kappa(\tau)$ is a fermionic (i.e. $\kappa^2 = 0$) transformation
parameter.  Furthermore, the transformation  $\delta_{sg}$ is a supersymmetric
transformation because it transforms a bosonic variable to a fermionic variable and 
{\it vice-versa}. The transformations in Eq. (2) are {\it symmetry} transformations because 
the first-order Lagrangian $L_f$ transforms to the following total derivatives:
\begin{eqnarray}
\delta_g \, L_f = \frac{d}{d\tau} \, \Big[\frac{\xi}{2}\,(p^2 + m^2)\Big], \qquad \qquad 
\delta_{sg} \, L_f = \frac{d}{d\tau} \, \Big[\frac{\kappa}{2}\,
(p_\mu \, \psi^\mu + m \, \psi_5)\Big]. 
\end{eqnarray}
As a consequence, it is clear that the action integral $S = \int_{- \infty}^{+ \infty} d\,\tau \, L_f$,
under the transformations $\delta_{g}$ and $\delta_{sg}$, would be equal to {\it zero} (i.e. $\delta_p S = 0, p = g, sg $) due to the fact
that all the {\it physical} variables vanish off at $\tau = \pm \, \infty$. 
There is a reparameterization symmetry, too, in our theory due to the basic infinitesimal transformation 
$\tau \rightarrow \tau^{'} = \tau - \epsilon(\tau)$ where $\epsilon(\tau)$ is an infinitesimal
transformation parameter. In fact, the physical variables of our 1D system transform under the
infinitesimal reparameterization transformation $(\delta_r)$ as:
\begin{eqnarray}
&&\delta_{r} \, x_\mu = \epsilon \, {\dot x}_\mu, \qquad \qquad \delta_{r} \, p_\mu = 
\epsilon \, {\dot p}_\mu, \qquad \qquad \delta_{r} \, \psi_\mu = \epsilon \, {\dot \psi}_\mu,\nonumber \\
&&\delta_{r} \,\psi_5 = \epsilon \, {\dot \psi}_5, \qquad \qquad  \delta_{r} \, e = \frac{d}{d\tau}(\epsilon\,e), \qquad \qquad 
\delta_{r} \, \chi = \frac{d}{d\tau}(\epsilon\,\chi).
\end{eqnarray}
The above transformations are {\it symmetry} transformations for the action integral $S = \int_{- \infty}^{+ \infty} d\,\tau \, L_f$
because of the following transformation property of $L_f$, namely;
\begin{eqnarray}
\delta_r \, L_f = \frac{d}{d \tau} \, \Big[\epsilon\,L_f \Big] \qquad  \implies \qquad \delta_{r} \, S = 0.
\end{eqnarray}
It is evident that $\delta_{r} \, S = 0$ due to the fact that $\epsilon(\tau)$ and $L_f$ vanish-off at $\tau = \pm \, \infty$.

The reparameterization symmetry transformation $(\delta_r)$ and gauge symmetry transformation
$(\delta_g)$ are {\it equivalent} under the following limits
\begin{eqnarray}
\xi = \epsilon \, e, \qquad \qquad  {\dot x}_\mu = e \, p_\mu, \qquad \qquad {\dot p}_\mu = 0,  
\end{eqnarray}
provided we set {\it all} the fermionic variables $(\chi, \psi_5, \psi_\mu)$ of our theory  
equal to {\it zero}. In the above, we have used equations of motion: 
${\dot x}_\mu = e \, p_\mu$ and ${\dot p}_\mu = 0$ {\it and} we have identified the gauge symmetry transformation parameter $\xi(\tau)$ with the 
combination of the reparameterization transformation parameter $\epsilon(\tau)$ and the  einbein variable $e(\tau)$. In exactly similar
fashion, we note that $(\delta_r)$ and $(\delta_g + \delta_{sg})$ are also {\it equivalent}.   
In this context, first of all, we note that there are two primary constraints (i.e. $\Pi_e \approx 0,\,
\Pi_\chi \approx 0)$ and two secondary constraints (i.e. $ p^2\,-\,m^2\approx \,0,\, p_\mu\, \psi^\mu - m\,
\psi_5\approx 0)$ on our theory where $\Pi_e$ and $\Pi_\chi$ are the canonical conjugate momenta w.r.t. the  
Lagrange  multiplier variables $e$ and $\chi$, respectively\footnote{The above {\it four} constraints of our theory  are
{\it first-class} in the terminology of Dirac's prescription for the classifications scheme of 
constraints because they (anti)commute among themselves 
[36, 37].}. These constraints generate the combined (super)gauge symmetry transformations 
$ \delta\, =\, \delta_g\,+\,\delta_{sg} $ for the physical variables of our theory as (see, e.g. [27])
\begin{eqnarray}
&&\delta\,x_\mu = \xi \, p_\mu +\kappa\, \psi_\mu, \qquad \delta p_\mu = 0,\qquad\delta\psi_\mu = i\,\kappa\,p_\mu,\nonumber\\
&&\delta e = \dot\xi +2\,\kappa\,\chi,\qquad\delta\chi  = i\,\dot\kappa,\qquad \delta\psi_5 = i\,\kappa\,m,
\end{eqnarray}
under which the first-order Lagrangian $L_f$ transforms to a total ``time" derivative as:
\begin{eqnarray}
\delta L_f  = \frac {d}{d\tau}\,\Big[\frac{\xi}{2}\,(p^2 + m^2) + \frac{\kappa}{2}\,(p_\mu\,\psi^\mu + m\, \psi_5)\Big].
\end{eqnarray}
As a consequence of the above observation, it is evident that $\delta S = 0$ where 
$S = \int_{-\infty}^{+\infty} d\tau\,L_f$ is the action integral. If we use the following equations of motion:
\begin{eqnarray}
\dot p_\mu = 0, \qquad \dot x_\mu = e\,p_\mu - i\,\chi\,\psi_\mu,\qquad \dot \psi_\mu = \chi\, p_\mu,\qquad \dot \psi_5 = m\,\chi, 
\end{eqnarray}
and identify the transformations parameters as 
\begin{eqnarray}
\xi = e\,\epsilon, \qquad\qquad \kappa = -\,i\, \epsilon\,\chi,
\end{eqnarray}
we find that the reparameterization symmetry transformation (4) [emerging due to the basic transformation:
$\tau \rightarrow \tau^{'} = \tau - \epsilon(\tau)$] {\it and} the combined gauge and supergauge symmetry 
transformations (i.e. $\delta = \delta_g + \delta_{sg}$), quoted in Eq. (7), 
are {\it equivalent} to each-other. It is worthwhile to note that, under the identifications (10), 
the transformation $\delta e = \dot\xi + 2\,\kappa\,\chi$ becomes $\delta e = \frac {d}{d\tau} (\epsilon\,e)$ 
as we note that $2\,\kappa\,\chi = -\,2\,i\,e\,\chi^2 = 0.$

We end this section with the following remarks. First of all, we note that the canonical Hamiltonians, 
derived from $L_0$ {\it and} $L_f$ (as well as $L_s$), are
\begin{eqnarray}
&& H_c^{(0)} = -\, i \, \chi \, (p_{\mu}\, \psi^{\mu} - m \, \psi_5), \qquad H_c = 
\frac{e}{2}\,(p^2 - m^2) -\, i \, \chi \, (p_{\mu}\, \psi^{\mu} - m \, \psi_5),
\end{eqnarray}
where $H_c^{(0)}$ is the canonical Hamiltonian corresponding to the Lagrangian $L_0$.
It is straightforward to note that the primary constraints $\Pi_{e} \approx 0$, $\Pi_{\chi} \approx 0$ lead to the derivation of the 
secondary constraints $(p^2 - m^2) \approx 0$, $(p_{\mu}\, \psi^{\mu} - m \, \psi_5) \approx 0$ from the Hamiltonians (11) as well as from 
all the {\it three} equivalent Lagrangians (1) (cf. Appendix A below). Second, we have explicitly
demonstrated that the (super)gauge symmetry transformations and reparameterization symmetry 
transformations are {\it equivalent} under {\it specific} conditions [cf. Eqs. (9), (10)]. Finally, the system under consideration 
is very interesting and important  because it is endowed with many symmetries and it provides
a prototype  example for the supersymmetric gauge theory, superstrings and a model for the supergravity theory.

\section {(Anti-)BRST Symmetries: Lagrangian Formulation}

Our present section is divided into two parts. In the subsection 3.1, we show the existence 
of the CF-type restriction by the requirement of absolute anticommutativity of the (anti-) BRST
symmetries and (anti-)BRST invariance of the coupled (but equivalent) Lagrangians  $L_b$ and
$L_{\bar b}$. In the subsection 3.2, we establish the existence of the {\it same} by requiring the absolute
anticommutativity of the conserved and nilpotent  (anti-)BRST charges.\\

\subsection{(Anti-)BRST Invariance  and CF-Type Restriction}

Corresponding to the combined {\it classical} (super)gauge symmetry transformations [cf. Eq. (7)],
we can write down the {\it quantum} (anti-)BRST symmetry transformations $s_{(a)b}$ where the classical gauge symmetry 
parameter $\xi (\tau)$ would be replaced by the fermionic ($c^2 = \bar c^2 = 0, c\,\bar c + \bar c\, c = 0, etc.$)
(anti-)ghost  variables $(\bar c)c$ and the {\it classical} supergauge symmetry transformations parameter  $\kappa (\tau)$
would be replaced by a pair of {\it bosonic}  $(\beta^2  = \bar\beta \neq 0)$ (anti-) ghost variables
$(\bar\beta)\beta$. These off-shell nilpotent $[(s_{(a)b})^2 = 0]$, infinitesimal and continuous (anti-)BRST symmetry 
transformations $(s_{(a)b})$, in their {\it full} blaze of glory for our 1D system of the {\it massive}  
spinning relativistic particle,  are (see, e.g. [25])
\begin{eqnarray} 
&& s_{ab}\; x_\mu = {\bar c}\; p_\mu + \bar \beta \;\psi_\mu, \quad\qquad s_{ab}\; e = \dot {\bar c} + 2 \;\bar \beta\; \chi,  
\;\quad\qquad s_{ab} \;\psi_\mu = i \;\bar \beta\; p_\mu,\nonumber\\
&& s_{ab}\; \bar c = - i \;{\bar \beta}^2, \quad s_{ab}\; c = i\; \bar b, \quad s_{ab}\; \bar \beta = 0, 
\;\quad s_{ab} \; \beta = - i\; \gamma, \quad s_{ab}\; p_\mu = 0, \nonumber\\
&& s_{ab} \;\gamma = 0, \quad s_{ab}\; \bar b = 0, \quad s_{ab}\;\chi = i\; \dot {\bar \beta}, 
\quad s_{ab} \; b =  2\; i\; \bar \beta\; \gamma,\quad s_{ab} \,\psi _5 = i\,\bar\beta\,m,
\end{eqnarray}
\begin{eqnarray}
&&s_b\; x_\mu = c\;p_\mu + \beta \;\psi_\mu, \quad\qquad s_b\; e = \dot c + 2\;\beta\; \chi,  
\quad\qquad s_b\; \psi_\mu = i\;\beta\; p_\mu,\nonumber\\
&& s_b\;c = - i\; \beta^2, \;\quad s_b \;{\bar c} = i\; b, \;\quad s_b \;\beta = 0, 
\;\quad s_b \;\bar \beta = i \;\gamma, \;\quad s_b\; p_\mu = 0,\nonumber\\
&& s_b \;\gamma = 0, \quad s_b \;b = 0, \quad s_b \;\chi = i\; \dot \beta, 
\qquad s_b\; \bar b = - 2\; i\; \beta\; \gamma,\quad s_{b}\, \psi _5 = i\,\beta\,m,
\end{eqnarray} 
where $b$ and $\bar b$ are the Nakanishi-Lautrup type auxiliary variables, fermionic ($\chi^2 = 0, c^2 = \bar c^2 = 0, \gamma^2  = 0$)
variables $(\chi, c, \bar c, \gamma)$ are present in our theory and rest of the symbols have already been explained earlier. 
As far as the absolute anticommutativity $(s_b\,s_{ab}  + s_{ab}\,s_b = 0)$ property is concerned, it can be checked that
\begin{eqnarray}
\{s_b, s_{ab}\}\,x_\mu = i\, (b + \bar b + 2\,\beta\bar\beta)\,p_\mu,\qquad
\{s_b, s_{ab}\}\,e  = i\, \frac{d}{d\tau}\, (b + \bar b + 2\,\beta\bar\beta),
\end{eqnarray}
are equal to zero {\it only} after imposing the CF-type restriction: $b + \bar b + 2\,\beta\bar\beta = 0$
from {\it outside}. It is worthwhile  to mention  that {\it this} CF-type restriction is a {\it physical} restriction
within the realm of BRST formalism because it is an (anti-)BRST invariant (i.e. $s_{(a)b}\, [b + \bar b + 2\,\beta\bar\beta] = 0)$
quantity. Except for the variables ($x_\mu, e$), it is straightforward to check that the following is true
for the {\it other} variables of our theory, namely; 
\begin{eqnarray}
\{s_b, s_{ab}\}\Phi  = 0, \qquad\qquad \Phi = p_\mu, \psi_\mu, \psi_5, \chi, \beta, \bar\beta, c, \bar c, b, \bar b, \gamma,
\end{eqnarray}
where $\Phi (\tau)$ is the {\it generic} variable of the (anti-)BRST invariant theory. Thus, it is crystal clear that 
the (anti-)BRST symmetry transformations in (12) and (13) are off-shell nilpotent $[(s_{(a)b})^2 = 0]$ and absolutely anticommuting
$(s_b \, s_{ab} + s_{ab} \, s_b = 0)$ in nature provided the whole theory is considered on a submanifold of space of {\it quantum} variables
where the CF-type restriction: $b + {\bar b} + 2\,\beta\,{\bar \beta} = 0$ is {\it satisfied} 
in the quantum Hilbert space (see, e.g. [25]).

The coupled (but equivalent) Lagrangians for our (anti-)BRST invariant system of the 
1D {\it massive} spinning relativistic particle can be written as:
\begin{eqnarray}
&& L_b = L_f + s_b \, s_{ab}\Big[\frac{i\,e^2}{2} + c\, {\bar c} + \chi\,\psi_5\Big], \nonumber \\
&& L_{\bar b} = L_f - s_{ab} \, s_b\Big[\frac{i\,e^2}{2} + c\, {\bar c} + \chi\,\psi_5\Big],
\end{eqnarray}
where $L_f$ is the first-order Lagrangian that has been quoted in Eq. (1). The above Lagrangians for our 1D system
of a massive spinning relativistic particle can be written, in their full glory incorporating the gauge-fixing 
and Faddeev-Popov ghost terms, as: 
\begin{eqnarray}
 L_b & = & L_f + b^2 +b\,({\dot e} + 2\, {\bar \beta}\,\beta) - i\,\dot{\bar c} \, {\dot c} + {\bar \beta}^2 \, {\beta}^2 
+ 2\, i\, \chi\,(\beta\,\dot{\bar c} - \bar{\beta} \, \dot c)  - 2\,e\,(\bar\beta \, \dot\beta + \gamma\,\chi) \nonumber\\
&+& 2\,\gamma\,(\beta \, \bar c - \bar\beta \, c) + m \, (\bar\beta \, \dot\beta - \dot{\bar \beta}\, \beta + \gamma \, \chi) 
- \dot \gamma \, \psi_5, 
\end{eqnarray}
\begin{eqnarray}
L_{\bar b} &=& L_f + {\bar b}^2 - \bar b\,({\dot e} - 2\, {\bar \beta}\,\beta) - i\,\dot{\bar c} \, {\dot c} + {\bar \beta}^2 \, {\beta}^2 
+ 2\, i\, \chi\,(\beta\,\dot{\bar c} - \bar{\beta} \, \dot c)  + 2\,e\,(\dot{\bar \beta} \, \beta - \gamma\,\chi) \nonumber\\
&+& 2\,\gamma\,(\beta \, \bar c - \bar\beta \, c) + m \, (\bar\beta \, \dot\beta - \dot{\bar \beta}\, \beta + \gamma \, \chi)
- \dot \gamma \, \psi_5,
\end{eqnarray}
where, as pointed out earlier, $b$ and $\bar b$ are the Nakanishi-Lautrup type  auxiliary variables which lead to the derivation of
EL-EOMs (from $L_b$ and $L_{\bar b}$) as:
\begin{eqnarray}
2\,b + \dot e +2\,\beta\,\bar\beta = 0,  \qquad 2\,{\bar b} - \dot e + 2\,\beta\,\bar\beta = 0. 
\end{eqnarray}
It is elementary to note that the above relationships lead to the derivation of the CF-type restriction: 
$b + \bar b + 2\, \bar\beta\,\beta = 0$ which is the hallmark of a quantum gauge theory discussed within
the framework of BRST formalism [5, 6].

At this juncture, we are in the position to focus on the symmetry properties of 
the coupled Lagrangians $L_b$ and $L_{\bar b}$. In this context, we observe the following:
\begin{eqnarray}
s_b \, L_b = \frac{d}{d\,\tau}\,\Big[\frac{c}{2}\,(p^2 + m^2) + \frac{\beta}{2}\,(p_\mu \, \psi^\mu + m \, \psi_5)
+ b \, (\dot c + 2\,\beta\, \chi)\Big],   
\end{eqnarray}
\begin{eqnarray}
s_{ab} \, L_{\bar b} = \frac{d}{d\,\tau}\,\Big[\frac{\bar c}{2}\,(p^2 + m^2) + \frac{\bar\beta}{2}\,(p_\mu \, \psi^\mu + m \, \psi_5) 
- \bar b \, (\dot{\bar c} + 2\,\bar\beta \, \chi)\Big].
\end{eqnarray}
It is clear from the above observations that the action integrals $S_1 = \int^\infty_{- \infty} d\,\tau\, L_b$ and 
$S_2 = \int^\infty_{- \infty} d\,\tau \,L_{\bar b}$ remain invariant (i.e. $s_b\,S_1 = 0, s_{ab}\,S_2 = 0$) under the {\it quantum} BRST and
anti-BRST symmetry transformations that have been listed in Eqs. (13) and (12). The coupled (but equivalent)
Lagrangian respect {\it both} (i.e. BRST and anti-BRST) {\it quantum} symmetries provided 
the whole theory is considered on a sub-manifold of the quantum Hilbert space of variables where
the CF-type restriction: $b+\bar b+2\beta \bar \beta=0$ is satisfied. In other words, mathematically, we observe the following:
\begin{eqnarray}
s_b \, L_{\bar b} & = & \frac{d}{d\,\tau}\,\Big[\frac{c}{2}\,(p^2 + m^2) + \frac{\beta}{2}\,(p_\mu \, \psi^\mu 
+ m \, \psi_5) - \bar b \, (\dot c + 2\,\beta \, \chi) + 2\,i\,e\,\beta \gamma\Big] \nonumber\\ 
& + & (\dot c + 2\beta \,\chi)\,\Big[\frac{d}{d\,\tau}\,(b+\bar b+2\beta \bar \beta)\Big]\,
- (2\,i\, \beta \gamma)\;(b+\bar b+2\beta \bar \beta), \end{eqnarray}
\begin{eqnarray}
s_{ab} \, L_{b} & = & \frac{d}{d\,\tau}\,\Big[\frac{\bar c}{2}\,(p^2 + m^2) + \frac{\bar \beta}{2}\,(p_\mu \, \psi^\mu 
+ m \, \psi_5) + b \, ( \dot{\bar c} + 2\,\bar \beta
\, \chi) + 2\,i\,e\,\bar \beta \gamma\Big] \nonumber\\ 
&-& (\dot{\bar c} + 2\bar \beta \,\chi)\,\Big[\frac{d}{d\,\tau}\,(b+\bar b+2\beta \bar \beta)\Big]\,
+ (2\,i\, \bar \beta \gamma) \;(b + \bar b + 2\beta \bar \beta).  
\end{eqnarray}
A close look at the above transformations demonstrates that if we impose the (anti-)BRST invariant
$ \big[s_{(a)b}\,(b+\bar b+2\,\beta\, \bar \beta)=0]$ {\it quantum} CF-type restriction $(b+\bar b+2\,\beta\, \bar \beta = 0)$
from {\it outside}, we obtain the following BRST symmetry transformation of
the Lagrangian $L_{\bar b}$ and anti-BRST symmetry transformation of the Lagrangian  $L_{b}$, namely;
\begin{eqnarray}
s_b \, L_{\bar b} = \frac{d}{d\,\tau}\,\Big[\frac{c}{2}\,(p^2 + m^2) + \frac{\beta}{2}\,(p_\mu \, \psi^\mu
+ m \, \psi_5) - \bar b \, (\dot c + 2\,\beta \, \chi) + 2\,i\,e\,\beta\, \gamma\Big],\\
s_{ab} \, L_{b} = \frac{d}{d\,\tau}\,\Big[\frac{\bar c}{2}\,(p^2 + m^2) + \frac{\bar \beta}{2}\,(p_\mu \, \psi^\mu + m \, \psi_5)
+ b \, ( \dot{\bar c} + 2\,\bar \beta\, \chi) + 2\,i\,e\,\bar \beta\, \gamma\Big].
\end{eqnarray}
It is crystal clear {\it now} that the observations in Eqs. (20), (21), (22), (23), (24) and  (25) imply,  in a 
straightforward manner, that {\it both} the Lagrangians (i.e. $L_{b}$ and $L_{\bar b}$) respect {\it both} the {\it quantum}
symmetries (i.e. BRST and anti-BRST symmetry transformations) in the space of {\it quantum}  variables where the CF-type
restriction is satisfied.

We end this sub-section with the following remarks. First and foremost,
we observe that the presence of the term ``$\chi\,\psi_5$" in the square-bracket of Eq. (16) is due to the
{\it massive} nature of the spinning relativistic particle. In the {\it massless} case, it disappears 
(see, e.g. Ref [25]). Second, the hallmark of the {\it quantum } gauge theory (within the framework of
the BRST formalism) is encoded in the existence of the CF-type restriction which we have demonstrated in
Eqs. (14), (19), (22) and (23) {\it where} we have concentrated on the {\it quantum} (anti-)BRST symmetries
which are respected by the coupled Lagrangians $L_b$ and $L_{\bar b}$. Finally, we note that the absolute 
anticommutativity property of the (anti-)BRST symmetries  and {\it equivalence} of  $L_b$ and $L_{\bar b}$ owe
their origins to the CF-type restriction: $b + \bar b + 2\, \beta\bar\beta = 0$.\\

\subsection{(Anti-)BRST Charges and CF-Type Restriction}

In this subsection, we demonstrate the existence of the (anti-)BRST invariant CF-type restriction (i.e.\,$b+\bar b+2\,\beta\, \bar \beta=0$)
by demanding the absolute anticommutativity of the conserved and nilpotent (anti-)BRST charges of our present theory. 
In this context, first of all,  we note that, according to Noether's theorem, the invariances $ s_b\, S_1 = 0,\,\, s_{ab}\, S_2 = 0$
of the action integrals  $S_1 = \int_{- \infty}^{+ \infty} d\,\tau \, L_b$ and  $S_2 = \int_{- \infty}^{+ \infty} d\,\tau
\, L_{\bar b}$ under the (anti-)BRST symmetry transformations ($s_{(a)b}$) [as quoted in Eqs. (20) and (21)] lead to the
derivation of the Noether conserved (anti-)BRST charges ($Q_{(a)b}^{(1)}$) as follows:
\begin{eqnarray}
Q_{ab}^{(1)}  &=& \frac{\bar c}{2}\,(p^2 - m^2) + {\bar \beta}\,(p_\mu \, \psi^\mu - m \, \psi_5)
- \bar b\,\dot{\bar c}  - 2\,\bar b\,\bar \beta
\, \chi - i\,m\,\bar \beta \gamma\,-\bar \beta{^2}\, \dot c- 2\,\beta\,\bar \beta{^2}\,\chi,\\
Q_{b}^{(1)}  &=& \frac{c}{2}\,(p^2 - m^2) + {\beta}\,(p_\mu \, \psi^\mu - m \, \psi_5)+
  b\,\dot c  + 2\, b\, \beta\, \chi - i\,m\, \beta \gamma\,+ \beta{^2}\, \dot {\bar c} + 2\,\bar \beta\, \beta{^2}\,\chi. 
\end{eqnarray}
The conservation law (i.e. $\dot Q_{b}^{(1)} = 0, \;\dot Q_{ab}^{(1)} = 0$) can be proven by using the EL-EOMs derived from the 
Lagrangians $ L_b$ and $L_{\bar b}$ (cf. Appendix B below). We have used the superscript (1) on the (anti-)BRST charges
($Q_{(a)b}^{(1)}$) to denote that these charges have been {\it directly} derived by using the basic principle behind  
Noether's theorem. However, we have the option of expressing these charges in a different 
form by using the EL-EOMs that are derived from $ L_b$ and $L_{\bar b}$. At this stage, it can
be noted that the Noether conserved charges $Q_{(a)b}^{(1)}$ are {\it not} off-shell nilpotent
($[Q_{(a)b}^{(1)}]^2 \neq 0$) of order two {\it without} any use of EL-EOMs. In other words, we note that
the following is true, namely;
\begin{eqnarray}
s_b \, Q_b^{(1)} &=& - \, i \, \{Q_b^{(1)}, Q_b^{(1)}\} \neq 0, \nonumber \\
s_{ab} \, Q_{ab}^{(1)} &=& - \, i \, \{Q_{ab}^{(1)}, Q_{ab}^{(1)}\} \neq 0,
\end{eqnarray}
{\it unless} we use the EL-EOMs from $L_b$ and $L_{\bar b}$. Thus, we lay emphasis on the 
fact that $Q_{(a)b}^{(1)}$ are {\it only} the {\it on-shell} nilpotent conserved charges  (even though we have used the off-shell nilpotent
(anti-)BRST symmetry transformations (12) and (13) in {\it their} derivation).

We have the freedom to use the EL-EOMs (derived from $L_{\bar b}$ and $L_b$) to recast the Noether conserved charges $Q_{(a)b}^{(1)}$
in a different form. For instance, the BRST charge $Q_b^{(1)}$ can be written in a {\it different} form by using the following
EL-EOMs
\begin{eqnarray}
&&{\dot b} = - \, \frac{1}{2} \, (p^2 - m^2) - 2 \, ({\bar \beta} \, {\dot \beta} + \gamma \, \chi), \nonumber \\
&&p_\mu \, \psi^\mu - m \, \psi_5 = 2 \, i \, e \, \gamma - 2 \, (\beta \, \dot{\bar c} - \bar \beta \, \dot c),
\end{eqnarray}
which are derived from $L_b$ w.r.t. the $e$ and $\chi$ variables. The ensuing expression for the conserved BRST  charge 
[due to EL-EOMs (29)] is:
\begin{eqnarray}
Q_b^{(2)} &=& b \, \dot c - \dot b \, c + 2 \, i \, e \, \beta \, \gamma + 2\,\beta\,\bar\beta\,\dot c
- 2 \, c \, (\bar \beta \, {\dot \beta} + \gamma \, \chi) + 2 \, b \, \beta \, \chi \nonumber \\ 
&-& 2\,i \, m \, \gamma \, \beta - \beta^2 \, \dot{\bar c}
+ 2 \, \chi \, \beta^2 \, \bar\beta. 
\end{eqnarray}
Here the superscript $(2)$ denotes that the expression for the BRST charge in Eq. (30) has been derived from the Noether conserved
BRST charge $Q_b^{(1)}$ by using the EL-EOMs quoted in Eq. (29). It is now straightforward to check that the following is true, namely;
\begin{eqnarray}
s_b \, Q_b^{(2)} = - \, i \, \{Q_b^{(2)}, Q_b^{(2)}\} = 0  \qquad \implies \qquad \big[Q_b^{(2)}\big]^2 = 0, 
\end{eqnarray}
where we have {\it directly} applied the BRST symmetry transformation (13) on the expression for $Q_b^{(2)}$ [cf. Eq. (30)] for the 
computation of the l.h.s. of Eq. (31). We would like to lay emphasis on the fact that Eq. (31) is nothing but the 
standard relationship between the continuous symmetry transformation $s_b$ and its generator $Q_b ^{(2)}$. The {\it latter} 
is, to be precise, the conserved BRST charge which is the generator of the symmetry transformations (13).
We, ultimately, note that the {\it off-shell} nilpotency $([Q_b^{(2)}]^2 = 0)$ of the $Q_b ^{(2)}$ has been proven in (31)
where we have {\it not} used any EL-EOMs and/or CF-type restriction.

Let us now concentrate on the proof of the off-shell nilpotency of the anti-BRST charge ($Q_{ab}$). For this purpose, we use the following
EL-EOMs 
\begin{eqnarray}
&&\dot{\bar b} = \frac{1}{2} \, (p^2 - m^2) - 2 \, (\dot{\bar \beta} \, { \beta} - \gamma \, \chi), \nonumber \\
&&p_\mu \, \psi^\mu - m \, \psi_5 = 2 \, i \, e \, \gamma - i \, m \, \gamma - 2 \, (\beta \, \dot{\bar c} - \bar \beta \, \dot c),
\end{eqnarray}
that emerge out from the Lagrangian $L_{\bar b}$ (when we consider the variables $e$ and $\chi$ for their derivation) to recast the
Noether conserved charge $Q_{ab}^{(1)}$ as
\begin{eqnarray}
Q_{ab}^{(2)} &=& \dot{\bar b} \, \bar c - \bar b \, \dot{\bar c} + 2 \, i \, e \, \bar\beta \, \gamma 
- 2 \, \beta \, \bar\beta \, \dot{\bar c}  + 2 \, \bar c \, (\dot{\bar \beta} \, \beta - \gamma \, \chi)
 - 2 \, \bar b \, \bar \beta \, \chi \nonumber \\ 
& - & 2 \, i \, m \, \bar\beta \, \gamma + {\bar\beta}^2 \, \dot c - 2 \, \beta \, {\bar \beta}^2 \, \chi, 
\end{eqnarray}
where the superscript (2) on the anti-BRST charge $Q_{ab}^{(2)}$ denotes the fact that {\it it} has been derived 
from the Noether conserved charge $Q_{ab}^{(1)}$.
We apply, at this stage, the anti-BRST symmetry transformations (12) {\it directly} on the anti-BRST charge $Q_{ab}^{(2)}$ to obtain:
\begin{eqnarray}
s_{ab} \, Q_{ab}^{(2)} = - \, i \, \{Q_{ab}^{(2)}, Q_{ab}^{(2)}\} = 0  \qquad \implies \qquad \big[Q_{ab}^{(2)}\big]^2 = 0. 
\end{eqnarray}
The above observation proves the off-shell nilpotency of the anti-BRST charge $Q_{ab}^{(2)}$ because we do {\it not} use 
EL-EOMs and/or CF-type restriction in {\it its} proof. In Eq. (34), we have used the basic
principle behind the continuous symmetries and their generators. There are other ways, too, to prove the off-shell
nilpotency ($[Q_{(a)b}^{(2)}]^2 = 0$) of the (anti-)BRST charges $Q_{(a)b}^{(2)}$. However, we have concentrated,
in our present endeavor, {\it only} on the  standard relationship between the continuous symmetries and their generators.

A couple of decisive features of the BRST formalism is the validity of the off-shell/on-shell nilpotency and absolute anticommutativity
properties  of the (anti-)BRST symmetries as well as the (anti-)BRST charges. We concentrate now on the proof of the absolute anticommutativity
of the conserved and nilpotent (anti-)BRST charges $Q_{(a)b}^{(2)}$. Toward this goal in mind, we first concentrate on the expression
for $Q_b^{(2)}$ [cf. Eq. (30)]. Applying {\it directly} the anti-BRST symmetry transformations (12) on it, we obtain the following:
\begin{eqnarray}
s_{ab} \, Q_{b}^{(2)} = i \, (b + {\bar b} + 2 \, \beta \, \bar\beta) \, \big[\dot{\bar b} + 2 \, \chi \, \gamma + 2 \, \beta \, 
\dot{\bar\beta}\big] - i \, {\bar b} \, \frac{d}{d \, \tau} \,\big[b + {\bar b} + 2 \, \beta \, \bar\beta\big].
\end{eqnarray}
In the terminology of the standard relationship between the continuous symmetry transformation $(s_{ab})$ and its generator $Q_{ab}^{(2)}$, it is
evident that the l.h.s. of Eq. (35) can be written in an explicit fashion as:
\begin{eqnarray}
s_{ab} \, Q_{b}^{(2)} = - \, i \, \{Q_{b}^{(2)}, Q_{ab}^{(2)}\}.
\end{eqnarray}
A close look at (35) and (36) demonstrates that the absolute anticommutativity of the conserved (anti-)BRST charges (that are off-shell
nilpotent of order two) is {\it true} if and only if the CF-restriction: $b + {\bar b} + 2 \, \beta \, \bar\beta = 0$ is imposed on the theory
from {\it outside}. However, as discussed earlier, this restriction, on the {\it quantum} theory, is a {\it physical} condition  because 
this CF-type restriction is an (anti-)BRST invariant quantity.

Let us now focus on the expression for the off-shell nilpotent $\big[(Q_{ab}^{(2)})^2 = 0\big]$ anti-BRST
charge $(Q_{ab}^{(2)})$ in Eq. (33). The direct application of the BRST symmetry transformation ($s_b$) of
Eq.(13) on the anti-BRST charge $ Q_{ab}^{(2)}$ in (33), yields the following:
\begin{eqnarray}
s_b\,Q_{ab}^{(2)} & = & i\,b\, \frac{d}{d\,\tau}\Big[b + \bar b + 2\,\beta\, \bar \beta \Big]\,
- i\,(b+\bar b+2\,\beta\, \bar \beta)\,\big[\dot b + 2\,\dot \beta\, \bar \beta+ 2\,\gamma\,\chi\big],
\end{eqnarray}
It is straightforward to note that the r.h.s. of (37) would be equal to zero if we {\it impose} 
the (anti-)BRST invariant CF-type restriction $(b+\bar b+2\,\beta\, \bar \beta =0)$ from {\it outside}.
Exploiting the beauty of the standard relationship between continuous symmetry transformation ($s_b$) 
and its generator (conserved and nilpotent BRST charge $Q_{b}^{(2)}$), we note that the l.h.s. of the above equation  can be written as
\begin{eqnarray}
s_{b} \, Q_{ab}^{(2)} & = & - \, i \, \{Q_{ab}^{(2)}, Q_{b}^{(2)}\} = 0,
\end{eqnarray}
provided, as stated earlier, we confine  ourselves on the sub-manifold of the {\it quantum} 
Hilbert space of variables where the CF-type restriction $(b+\bar b+2\,\beta\, \bar \beta=0)$ is satisfied.

We end this subsection with the following remarks. First and foremost, the existence of CF-type restriction
is the hallmark\footnote{We have been able to establish an intimate connection between the CF-type restriction and the geometrical objects called gerbes  [5, 6].
The existence of {\it this} restriction provides an {\it independent} identity to the BRST and anti-BRST symmetries 
and the corresponding (anti-)BRST charges.} of a {\it quantum} theory described within the framework of BRST formalism [5, 6]. Second, the 
CF-type restriction is responsible for the existence of the coupled (but equivalent) Lagrangians $L_b$ and
$L_{\bar b}$. Third, the absolute anticommutativity of the (anti-)BRST symmetries and corresponding (anti-)BRST
charges owe their origins to the CF-type restriction. Finally, we have been able to show that  $L_b$ and 
$L_{\bar b}$ {\it both} respect {\it both} the (anti-)BRST symmetries due to the existence of CF-type restriction.\\

\section {BRST Symmetry Transformations: ACSA}

We exploit the basic tenets of ACSA to BRST formalism to derive the proper off-shell nilpotent BRST symmetry transformation\footnote{It will
be noted that {\it only} the BRST symmetry transformations have been mentioned in Ref. [33] for the  spinning relativistic particle. However,
the {\it full} set of 
(anti-)BRST symmetry transformations and the corresponding (anti-)BRST invariant  CF-type restriction have been derived in our earlier work [25].}
(13) where we take into account the {\it anti-chiral} supervariables [defined on the (1, 1)-dimensional {\it anti-chiral} super sub-manifold
of the general (1, 2)-dimensional supermanifold]. The above {\it anti-chiral} supervariables are the generalizations of the {\it ordinary}
variables of Lagrangian $L_b$ and ${\bar b} (\tau)$ as follows:
\begin{eqnarray*}
x_\mu(\tau) \quad &\longrightarrow& \quad X_\mu(\tau, \bar\theta) = x_\mu(\tau) + \bar\theta \, R_\mu^{(1)}(\tau), \nonumber \\
p_\mu(\tau) \quad &\longrightarrow& \quad P_\mu(\tau, \bar\theta) = p_\mu(\tau) + \bar\theta \, R_\mu^{(2)}(\tau), \nonumber \\
e(\tau) \quad &\longrightarrow& \quad E(\tau, \bar\theta) = e(\tau) + \bar\theta \, f_1(\tau), \nonumber \\
\end{eqnarray*}
\begin{eqnarray}
c(\tau) \quad &\longrightarrow& \quad F(\tau, \bar\theta) = c(\tau) + \bar\theta \, b_1(\tau), \nonumber \\
{\bar c}(\tau) \quad &\longrightarrow& \quad {\bar F}(\tau, \bar\theta) = {\bar c}(\tau) + \bar\theta \, b_2(\tau), \nonumber \\
\beta(\tau) \quad &\longrightarrow& \quad \tilde{\beta}(\tau, \bar\theta) = {\beta}(\tau) + \bar\theta \, f_2(\tau), \nonumber \\
\bar\beta(\tau) \quad &\longrightarrow& \quad \tilde{\bar\beta}(\tau, \bar\theta) = \bar{\beta}(\tau) + \bar\theta \, f_3(\tau), \nonumber \\
\psi_\mu(\tau) \quad &\longrightarrow& \quad \Psi_\mu(\tau, \bar\theta) = {\psi}_\mu(\tau) + \bar\theta \, b_3(\tau), \nonumber \\
\psi_5(\tau) \quad &\longrightarrow& \quad  \Psi_5(\tau, \bar\theta) = {\psi}_5(\tau) + \bar\theta \, b_4(\tau), \nonumber \\
\chi(\tau) \quad &\longrightarrow& \quad \tilde{\chi}(\tau, \bar\theta) = {\chi}(\tau) + \bar\theta \, b_5(\tau), \nonumber \\
\gamma(\tau) \quad &\longrightarrow& \quad {\Gamma}(\tau, \bar\theta) = {\gamma}(\tau) + \bar\theta \, b_6(\tau), \nonumber \\
b(\tau) \quad &\longrightarrow& \quad B(\tau, \bar\theta) = b(\tau) + \bar\theta \, f_4(\tau), \nonumber \\
{\bar b}(\tau) \quad &\longrightarrow& \quad {\bar B}(\tau, \bar\theta) = {\bar b}(\tau) + \bar\theta \, f_5(\tau). 
\end{eqnarray}
In the above, we have taken the super expansions along the Grassmannian  ${\bar\theta}$-direction of the 
{\it anti-chiral} $(1, 1)$-dimensional super sub-manifold which is 
parameterized by the superspace coordinates $(\tau,\bar \theta)$. We note that, in the above super expansions,
the secondary variables $( R_{\mu}^{(1)},\, R_{\mu}^{(2)},\, f_1,\,f_2,\,f_3,\,f_4,\,f_5  )$ are 
{\it fermionic} and rest of the secondary variables $(b_1,\,b_2,\,b_3,\,b_4,\,b_5,\,b_6 )$ are {\it bosonic} in nature
due to the fermionic $(\bar \theta ^2 =0)$ nature of the Grassmannian variable $\bar \theta$.
It is elementary to state that, in the limit $\bar\theta = 0$, we retrieve {\it ordinary} variables of our theory
described by the Lagrangian $L_b$ and $\bar b (\tau)$.

The {\it trivial} BRST invariant quantities: $ s_b p_\mu=0,\, s_b \gamma=0,\, s_b \beta=0,\, s_b b=0 $ 
imply that the secondary variables $ R_{\mu}^{(2)} = b_6 = f_2 = f_4 =0 $. This is due to the fact that the  
{\it basic tenets} of ACSA requires that the BRST invariant quantities should be {\it independent} of the
Grassmannian variable $\bar \theta$ (which is a mathematical artifact in the superspace formalism).
In other words, we have the following
\begin{eqnarray}
&&P_{\mu}^{(b)} (\tau, \theta)  =  p_\mu (\tau) + \bar\theta\, (0) \equiv p_\mu (\tau) + \bar\theta\, (s_b p_\mu (\tau))
\Longrightarrow P_{\mu}^{(b)} (\tau, \theta)  =  p_\mu (\tau), \nonumber \\
&&\Gamma^{(b)} (\tau, \theta)  =  \gamma (\tau) + \bar\theta\, (0)\;\, \equiv \gamma (\tau) + \bar\theta\, (s_b \gamma (\tau))\;\;
\Longrightarrow \Gamma ^{(b)} (\tau, \theta)  =  \gamma (\tau), \nonumber \\
&&B^{(b)} (\tau, \theta)  =  b (\tau) + \bar\theta\, (0) \;\,\equiv b (\tau) + \bar\theta\, (s_b b (\tau))\;\;\;
\Longrightarrow B ^{(b)} (\tau, \theta)  =  b (\tau), \nonumber \\
&&{\tilde \beta}^{(b)} (\tau, \theta)  =  \beta (\tau) + \bar\theta\, (0)\;\, \equiv \beta (\tau) + \bar\theta\, (s_b \beta (\tau))\;\;
\Longrightarrow {\tilde \beta} ^{(b)} (\tau, \theta)  =  \beta (\tau), 
\end{eqnarray}
where the superscript $(b)$ on the anti-chiral supervariables denotes the supervariables that
have been obtained after the application of the BRST invariant $ (s_b p_\mu =  s_b \gamma=  s_b b = s_b \beta = 0)$
restrictions so that the coefficients of $\bar \theta$, in the expansions (39),
becomes zero. This is due to the fact that there is a mapping (i.e. $s_b\leftrightarrow \partial_{\bar\theta},\;
s_{ab}\leftrightarrow \partial_\theta$) between the (anti-) BRST symmetry transformations $(s_{(a)b})$ and the translational  
operators $(\partial_\theta, \partial_{\bar\theta})$ along the Grassmannian directions of the (1, 2)-dimensional 
supermanifold that has been established in Refs. [10-12]. It is crystal clear, from our discussions in this paragraph, that
we have to determine precisely {\it all} the secondary variables in terms of the basic and auxiliary
variables of our theory so that we could know the coefficients of $\bar\theta$ in the super expansions (39).

Against the backdrop of our earlier  discussions, we have to obtain the precise expressions for the secondary variables 
so that we could obtain the BRST symmetry transformations $(s_b)$ as the coefficient of $\bar\theta$
in the {\it anti-chiral} super expansions (39). Toward this goal in our mind, we have to find out the {\it specific} combinations of the
{\it non-trivial} quantities that are BRST invariant. In this context, we note that the following useful and interesting quantities are BRST
invariant, namely;
\begin{eqnarray}
&&s_b(\bar\beta\;\gamma) = 0, \qquad\qquad\;\; s_b(\dot c + 2\,\beta\chi) = 0,\qquad \quad 
s_b( e\,\gamma\,\chi + e\,\bar\beta\,\dot\beta - i\;\bar\beta\,\dot c\, \chi) = 0,\nonumber\\
&& s_b(\beta^2 \bar\beta  + c\,\gamma) = 0,\quad\quad s_b(c\,p_\mu + \beta\psi_\mu) = 0, \qquad 
s_b(\beta\,x_\mu - i\, c\,\psi_\mu) = 0,\nonumber\\
&& s_b (b\,\bar\beta + \gamma\,\bar c) = 0, \;\;\qquad s_b(\bar b + 2\,\beta\bar\beta) 
= 0,\qquad\;\quad s_b (\dot \psi_5 - \chi\,m) = 0.
\end{eqnarray}
The basic tenets of ACSA to BRST formalism requires that the above quantities, at the {\it quantum} level,
should be {\it independent} of the Grassmannian variable 
$(\bar\theta)$ when {\it these} are generalized onto the $(1, 1)$-dimensional {\it anti-chiral} super sub-manifold of the general
$(1, 2)$-dimensional supermanifold. As a consequence, we have the following restrictions on the specific combinations of the {\it anti-chiral}
supervariables, namely;
\begin{eqnarray}
&&  \tilde{\bar\beta} (\tau, \bar\theta)\,\Gamma ^{(b)} (\tau, \bar\theta) = \bar\beta (\tau)\;\gamma(\tau),
\quad \dot F (\tau, \bar\theta) + 2\,{\tilde\beta}^{(b)} (\tau, \bar\theta)\,\tilde{\chi} (\tau, \bar\theta) 
= \dot c (\tau) + 2\,\beta (\tau)\, \chi (\tau),\nonumber\\
&& E (\tau, \bar\theta)\,\Gamma ^{(b)} (\tau, \bar\theta)\,\tilde{\chi} (\tau, \bar\theta) + 
E(\tau, \bar\theta)\,\tilde{\bar\beta} (\tau, \bar\theta)\,\dot {\tilde\beta}^{(b)} (\tau, \bar\theta)  
- i\, \tilde{\bar\beta} (\tau, \bar\theta)\,\dot F (\tau, \bar\theta)\,\tilde\chi (\tau, \bar\theta)\nonumber\\
&& =  e (\tau)\,\gamma (\tau)\,\chi (\tau) + e (\tau)\,\bar\beta (\tau)\,\dot\beta (\tau) 
- i\,\bar\beta (\tau)\,\dot c (\tau)\, \chi (\tau),\nonumber\\
&& \tilde{\beta} ^ {2(b)} (\tau, \bar\theta) \, \tilde{\bar\beta} (\tau, \bar\theta) + F (\tau, \bar\theta)\,\Gamma ^{(b)} (\tau, \bar\theta)
= \beta^2 (\tau)\, \bar\beta (\tau) + c(\tau)\,\gamma (\tau),\nonumber\\
&& F (\tau, \bar\theta)\,P_\mu ^{(b)} (\tau, \bar\theta) + \tilde{\beta} ^ {(b)} (\tau, \bar\theta)\,\Psi_\mu (\tau, \bar\theta)
= c(\tau)\,p_\mu (\tau) + \beta(\tau)\,\psi_\mu (\tau), \nonumber\\
&& \tilde{\beta} ^ {(b)} (\tau, \bar\theta)\,X_\mu  (\tau, \bar\theta) - i\, F (\tau, \bar\theta) \,\Psi_\mu (\tau, \bar\theta) 
= \beta (\tau)\,x_\mu (\tau)- i\, c (\tau)\,\psi_\mu (\tau),\nonumber\\
&& B^{(b)} (\tau, \bar\theta)\,\tilde{\bar\beta} (\tau, \bar\theta) + \Gamma ^{(b)} (\tau, \bar\theta)\, \bar F (\tau, \bar\theta)
= b (\tau)\,\bar\beta (\tau) + \gamma (\tau)\,\bar c (\tau),\nonumber\\
&&\bar B (\tau, \bar\theta) + 2\,\tilde{\beta} ^ {(b)} (\tau, \bar\theta)\, \tilde{\bar\beta} (\tau, \bar\theta) 
= \bar b (\tau) + 2\,\beta (\tau)\, \bar\beta (\tau),\nonumber\\
&& \dot \Psi_5 (\tau, \bar\theta) - \tilde \chi (\tau, \bar\theta)\,m = \dot \psi_5 (\tau) - \chi (\tau)\,m.
\end{eqnarray}
The above  restrictions are {\it quantum} gauge (i.e. BRST) invariant conditions on the {\it anti-chiral} supervariables
where the supervariables with superscript $(b)$ have been derived and explained in Eq. (40) that corresponds to the
{\it trivial} BRST symmetry transformations.

The substitutions of the {\it anti-chiral} super expansions (39) and the {\it trivial} expansions (40) 
into (42) lead to the following precise expressions for the secondary variables in terms of the {\it basic}
and {\it auxiliary} variables of the coupled (but equivalent) (anti-)BRST invariant Lagrangians $L_b$ and $L_{\bar b}$ [cf. Eqs. (17), (18)], namely;
\begin{eqnarray}
&&R_{\mu}^{(1)} = c\,p_\mu + \beta \,\psi_\mu,\quad\,  f_1 = \dot c + 2\;\beta\; \chi,\quad \,  
 b_1 = i\; \beta^2, \quad \,  b_2 = i\; b,\nonumber\\
 &&f_3 = i\;\gamma,\quad  b_3 = i\;\beta\; p_\mu,\quad  b_4 = i\,\beta\,m, \quad b_5 = i\; \dot \beta, \quad
 f_5 = - 2\; i\; \beta\;\gamma.
 \end{eqnarray}
Ultimately, we obtain the super expansions of (39) in terms of the off-shell nilpotent ($s_b ^2 = 0$) BRST transformations (13) of our theory  as follows
\begin{eqnarray*}
&& X_{\mu}^{(b)} (\tau,\bar \theta) = x_\mu + \bar\theta\,(c\,p_\mu + \beta \,\psi_\mu)\,\equiv x_\mu\,(\tau) 
+ \bar\theta \, (s_b x_\mu),\nonumber\\
&& E^{(b)} (\tau, \bar\theta) = e(\tau) + \bar\theta \, (\dot c + 2\;\beta\; \chi) \equiv e(\tau) + \bar\theta \, (s_b e), \nonumber \\
&& F^{(b)}(\tau, \bar\theta) = c(\tau) + \bar\theta \, (i\; \beta^2) \equiv c(\tau) + \bar\theta \, (s_b c), \nonumber \\
&& {\bar F}^{(b)}(\tau, \bar\theta) = {\bar c}(\tau) + \bar\theta \, (i\,b) \equiv {\bar c}(\tau) + \bar\theta \, (s_b \bar c), \nonumber \\
&& \tilde{\bar\beta}^{(b)}(\tau, \bar\theta) = \bar{\beta}(\tau) + \bar\theta \, (i\;\gamma) \equiv \bar{\beta}(\tau)
+ \bar\theta \, (s_b {\bar\beta}), \nonumber \\
&& \Psi_\mu^{(b)}(\tau, \bar\theta) = {\psi}_\mu(\tau) + \bar\theta \, (i\;\beta\; p_\mu) \equiv {\psi}_\mu(\tau) 
+ \bar\theta \, (s_b \psi_\mu), \nonumber \\
\end{eqnarray*}
\begin{eqnarray}
&& \Psi_5^{(b)}(\tau, \bar\theta) = {\psi}_5(\tau) + \bar\theta \, (i\,\beta\,m) \equiv {\psi}_5(\tau) + \bar\theta \, (s_b \psi_5), \nonumber \\
&& \tilde{\chi}^{(b)}(\tau, \bar\theta) = {\chi}(\tau) + \bar\theta \, (i\; \dot \beta) \equiv {\chi}(\tau) 
+ \bar\theta \, (s_b \chi), \nonumber \\
&&{\bar B}^{(b)}(\tau, \bar\theta) = {\bar b}(\tau) + \bar\theta \, (- \,2\; i\; \beta\;\gamma) 
\equiv {\bar b}(\tau) + \bar\theta \, (s_b {\bar b}),
\end{eqnarray}
which are {\it besides} the super expansions in (40) [that determine the {\it trivial} BRST symmetry transformations 
as: $ s_b\, p_\mu = 0,\, s_b\, \gamma = 0,\, s_b \,\beta = 0,\, s_b\,  b = 0$]. The superscript $(b)$ on the 
{\it anti-chiral} supervariable on the l.h.s. of the above expansions  denotes the fact that {\it these}
supervariables have been determined {\it after} the {\it quantum} gauge\footnote{At the {\it classical} level, we know that the gauge invariant 
quantities (GIRs) are {\it physical} objects. Within the framework  of BRST formalism, {\it all} the (anti-)BRST invariant 
quantities are {\it physical} objects at the {\it quantum} level. Hence, {\it these} quantities should be {\it independent} of the Grassmannian variables 
$(\theta, \bar\theta)$. In fact, this requirement is one of the basic tenets of ACSA to BRST formalism which is quite {\it physical}.} (i.e. BRST) invariant restrictions have been imposed on the supervariables as quoted in Eq. (42). 
In our Appendix C, we collect the step-by-step  computations that lead to the derivation of (43) from (42).

We end this section with the following remarks. First of all, we note that the coefficients of $\bar \theta$
in the super expansions (40) and (44) are nothing but the BRST transformations (13). Second, it is evident that $\partial_{\bar \theta}\;\Omega^{(b)}(x,\bar \theta) = s_b\,\omega (\tau)$
where $ \Omega^{(b)}(x,\bar \theta)$ is the generic  {\it anti-chiral} supervariable that is located on the 
l.h.s. of Eqs. (40) and (44) {\it and} the symbol $\omega (\tau)$ corresponds to the  generic  {\it ordinary} variable that is present in 
the Lagrangians $L_b$ and $L_{\bar b}$. Finally, we observe that, due to the mapping $s_b \leftrightarrow \partial_ {\bar \theta}$,
the off-shell nilpotency $(s_b^2=0)$ of the BRST symmetry transformations (13) is deeply connected with the nilpotency 
$( \partial_{\bar\theta}^{2} = 0)$ of the translational generator $(\partial_{\bar \theta})$ along the $\bar \theta$-direction 
of (1, 1)-dimensional {\it anti-chiral} super sub-manifold on which the {\it anti-chiral} supervariables are defined.\\

\section{Anti-BRST Symmetry Transformations: ACSA}

In this section, we derive the anti-BRST symmetry transformations (12) by exploiting the theoretical
potential and  power of  ACSA to BRST formalism. Toward this objective in mind, first of all, we 
generalize the {\it ordinary } variables of $L_{\bar b}$ [and the auxiliary variable $b(\tau)$] onto (1, 1)-dimensional
{\it chiral} super sub-manifold of the {\it general} (1, 2)-dimensional supermanifold (on which our 1D {\it ordinary} theory is generalized) as:
\begin{eqnarray*}
x_\mu(\tau) \quad &\longrightarrow& \quad X_\mu(\tau, \theta) = x_\mu(\tau) + \theta \, {\bar R}_\mu^{(1)}(\tau), \nonumber \\
p_\mu(\tau) \quad &\longrightarrow& \quad P_\mu(\tau, \theta) = p_\mu(\tau) + \theta \, {\bar R}_\mu^{(2)}(\tau), \nonumber \\
e(\tau) \quad &\longrightarrow& \quad E(\tau, \theta) = e(\tau) + \theta \, {\bar f}_1(\tau), \nonumber \\
c(\tau) \quad &\longrightarrow& \quad F(\tau, \theta) = c(\tau) + i\,\theta \, {\bar b}_1(\tau), \nonumber \\
{\bar c}(\tau) \quad &\longrightarrow& \quad {\bar F}(\tau, \theta) = {\bar c}(\tau) + i\,\theta \, {\bar b}_2(\tau), \nonumber \\
\beta(\tau) \quad &\longrightarrow& \quad \tilde{\beta}(\tau, \theta) = {\beta}(\tau) + \theta \, {\bar f}_2(\tau), \nonumber \\
\bar\beta(\tau) \quad &\longrightarrow& \quad \tilde{\bar\beta}(\tau, \theta) = \bar{\beta}(\tau) + \theta \, {\bar f}_3(\tau), \nonumber \\
\psi_\mu(\tau) \quad &\longrightarrow& \quad \Psi_\mu(\tau, \theta) = {\psi}_\mu(\tau) + \theta \, {\bar b}_3(\tau), \nonumber \\
\psi_5(\tau) \quad &\longrightarrow& \quad  \Psi_5(\tau, \theta) = {\psi}_5(\tau) + \theta \, {\bar b}_4(\tau), \nonumber \\
\end{eqnarray*}
\begin{eqnarray}
\chi(\tau) \quad &\longrightarrow& \quad \tilde{\chi}(\tau, \theta) = {\chi}(\tau) + \theta \, {\bar b}_5(\tau), \nonumber \\
\gamma(\tau) \quad &\longrightarrow& \quad {\Sigma}(\tau, \theta) = {\gamma}(\tau) + \theta \, {\bar b}_6(\tau), \nonumber \\
b(\tau) \quad &\longrightarrow& \quad B(\tau, \theta) = b(\tau) + \theta \, {\bar f}_4(\tau), \nonumber \\
{\bar b}(\tau) \quad &\longrightarrow& \quad {\bar B}(\tau, \theta) = {\bar b}(\tau) + \theta \, {\bar f}_5(\tau), 
\end{eqnarray}
where the $(1, 1)$-dimensional {\it chiral} super sub-manifold is parameterized by the superspace coordinates $(\tau, \theta)$ and
{\it all} the {\it chiral} supervariables on the l.h.s. of (45) are function of these superspace coordinates. The fermionic $(\theta^2 = 0)$
nature of the Grassmannian variable $\theta$ implies that the secondary variables $(\bar R_\mu^{(1)}, \bar R_\mu^{(2)}, {\bar f}_1, {\bar f}_2, 
{\bar f}_3, {\bar f}_4, {\bar f}_5)$ are {\it fermionic} and $(\bar b_1, \bar b_2, \bar b_3, \bar b_4, \bar b_5,
\bar b_6)$ are {\it bosonic} in nature. It is straightforward to note that, in the limit $\theta = 0$,
we retrieve our {\it ordinary} variables of Lagrangian $L_{\bar b}$ and the variable $b(\tau)$.

We note that there are {\it trivially} anti-BRST invariant quantities [cf. Eq. (12)] such as: $s_{ab}\,p_\mu = 0,\, s_{ab}\,
\gamma = 0,\, s_{ab}\,{\bar b} = 0, \,s_{ab}\,\bar\beta = 0$. As a consequence, we have the following 
{\it trivial chiral} super expansions (with inputs: $\bar R_\mu ^{(2)} = \bar b_6 = \bar f_5 = \bar f_3 = 0$), namely;
\begin{eqnarray}
&&p_\mu \, (\tau) \longrightarrow  P_\mu ^{(ab)}\, (\tau,\theta) = p_\mu (\tau) + \theta \, (0) 
\equiv p_\mu (\tau) + \theta \, (s_{ab} \,p_\mu (\tau)) ,\nonumber\\
&&\gamma (\tau) \longrightarrow  \Gamma ^{(ab)}\, (\tau,\theta) = \gamma (\tau) + \theta \, (0) 
\equiv \gamma (\tau) + \theta \, (s_{ab} \,\gamma (\tau)) ,\nonumber\\
&&\bar b (\tau) \longrightarrow  \bar B ^{(ab)}\, (\tau,\theta) = \bar b (\tau) + \theta \, (0) 
\equiv \bar b (\tau) + \theta \, (s_{ab}\, \bar b (\tau)) ,\nonumber\\
&&\bar \beta (\tau) \longrightarrow  \tilde {\bar \beta} ^{(ab)}\, (\tau,\theta) = \bar \beta (\tau) + \theta \, (0) 
\equiv \bar \beta (\tau) + \theta \, (s_{ab}\, \bar\beta (\tau)) ,
\end{eqnarray}
where the superscript $(ab)$ on the supervariables denotes the {\it chiral} supervariables where the coefficient 
of $\theta$ yields the anti-BRST symmetry transformations (12) in view the of mapping: $s_{ab} \leftrightarrow \partial_ {\theta}$ [10-12]
which becomes transparent when we observe that $\partial_\theta \,\Omega^{(ab)} (\tau,\theta) =  s_{ab} \, \omega (\tau)$
for the generic {\it supervariable} $\Omega^{(ab)} (\tau,\theta)$ and the corresponding {\it ordinary} generic variable 
$\omega (\tau)$. The trivial super expansion (46) would be utilized in our further discussions.

The basic ingredient of the ACSA to BRST formalism requires that the non-trivial anti-BRST invariant
quantities must be {\it independent} of the Grassmannian variable $\theta$ when {\it these} quantities
are generalized onto the (1, 1)-dimensional {\it chiral} super sub-manifold. We exploit {\it this} idea to 
determine the {\it secondary} variables of the super expansion (45) in terms of the {\it basic} and {\it  auxiliary}
variables of $L_{\bar b}$. Toward this aim in our mind, we note that the following anti-BRST invariant quantities
\begin{eqnarray}
&&s_{ab} (\beta\;\gamma) = 0, \qquad\qquad\;\; s_ {ab} (\dot {\bar c} + 2\,\bar\beta\chi) = 
0,\qquad \quad s_ {ab} ( e\,\gamma\,\chi - e\,\dot{\bar\beta}\,\beta + i\;\beta\,\dot {\bar c}\, \chi) = 0,\nonumber\\
&& s_ {ab} (\beta \bar\beta^2  - \bar c\,\gamma) = 0,\quad\quad s_{ab} (\bar c\,p_\mu + \bar\beta\psi_\mu) =
0, \qquad s_ {ab} (\bar\beta\,x_\mu - i\, \bar c\,\psi_\mu) = 0,\nonumber\\
&& s_{ab}  (\bar b\,\beta - \gamma\,c) = 0, \;\;\qquad s_{ab} (b + 2\,\bar\beta\beta) = 
0,\qquad\;\quad s_{ab} (\dot \psi_5 - \chi\,m) = 0, 
\end{eqnarray}
are found to be very useful and interesting because {\it their} generalizations
onto the (1, 1)-dimensional {\it chiral} super sub-manifold, namely;
\begin{eqnarray*}
&&  \tilde{\beta} (\tau, \theta)\,\Gamma ^{(ab)} (\tau, \theta) = \beta (\tau)\;\gamma(\tau),\quad \dot {\bar F} (\tau, \theta)
+ 2\,\tilde{\bar\beta}^{(ab)} (\tau, \theta)\,\tilde{\chi} (\tau, \theta) = \dot {\bar c} (\tau)
+ 2\,\bar\beta (\tau)\, \chi (\tau),\nonumber\\
&& E (\tau, \theta)\,\Gamma ^{(ab)} (\tau, \theta)\,\tilde{\chi} (\tau, \theta) 
- E(\tau, \theta)\,\dot{\tilde{\bar\beta}} (\tau, \theta)\,{\tilde\beta}^{(ab)} (\tau, \theta)  + i\, \tilde{\beta} (\tau, \theta)\,
\dot {\bar F} (\tau, \theta)\,\tilde\chi (\tau, \theta)\nonumber\\
&& =  e (\tau)\,\gamma (\tau)\,\chi (\tau) - e (\tau)\,\dot{\bar\beta} (\tau)\,\beta (\tau)
+ i\,\beta (\tau)\,\dot {\bar c} (\tau)\, \chi (\tau),\nonumber\\
&& \tilde{\beta}  (\tau, \theta) \, \tilde{\bar\beta}^ {2(ab)} (\tau, \theta) - \bar F (\tau, \theta)\,\Gamma ^{(ab)} (\tau, \theta) 
= \beta (\tau)\, \bar\beta ^ {(2)} (\tau) -  \bar  c(\tau)\,\gamma (\tau),
\nonumber\\
\end{eqnarray*}
\begin{eqnarray}
&&\bar F (\tau, \theta)\,P_\mu ^{(ab)} (\tau, \theta) + \tilde{\bar\beta} ^ {(ab)} (\tau, \theta)\,\Psi_\mu (\tau, \theta) 
= \bar c(\tau)\,p_\mu (\tau) + \bar\beta(\tau)\,\psi_\mu (\tau), \nonumber\\
&& \tilde{\bar\beta} ^ {(ab)} (\tau, \theta)\,X_\mu  (\tau, \theta) - i\, \bar F (\tau, \theta) \,\Psi_\mu (\tau, \theta)
= \bar\beta (\tau)\,x_\mu (\tau)- i\, \bar c (\tau)\,\psi_\mu (\tau),\nonumber\\
&& \bar B^{(ab)} (\tau, \theta)\,\tilde{\beta} (\tau, \theta) - \Gamma ^{(ab)} (\tau, \theta)\, F (\tau, \theta) 
= \bar b (\tau)\,\beta (\tau) - \gamma (\tau)\,c (\tau),\nonumber\\
&&B (\tau, \theta) + 2\,\tilde{\bar\beta}^ {(ab)} (\tau, \theta)\;\tilde{\beta}  (\tau, \theta) 
= b (\tau) + 2\,\bar\beta (\tau)\, \beta (\tau),\nonumber\\
&& \dot \Psi_5 (\tau, \theta) - \tilde \chi (\tau, \theta)\,m = \dot \psi_5 (\tau) - \chi (\tau)\,m,
\end{eqnarray}
yield the precise values of the secondary variables of the expansion in (45).
To be more precise, we note that the equalities  in (48) lead to:
\begin{eqnarray}
&&{\bar R}_\mu^{(1)} = {\bar c}\,p_\mu + \bar\beta \, {\psi_\mu}, \qquad {\bar f}_1 = \dot{\bar c} + 2\,\bar\beta\,\chi, \qquad 
{\bar b}_1 = i \, \bar b, \qquad {\bar b}_2 = -\,i\,{\bar\beta}^2, \nonumber \\ 
 &&{\bar f}_2 = -\,i\,\gamma, \qquad {\bar b}_3 = i \,\bar\beta\,p_\mu, 
\qquad {\bar b}_4 = i\,\bar\beta\,m, \qquad {\bar b}_5 = i\,\dot{\bar\beta}, \qquad {\bar f}_4 = 2 \, i \,\bar\beta\,\gamma.
\end{eqnarray}
Thus, we have determined precisely the expressions for the secondary variables in terms of the {\it basic} and {\it auxiliary}
variables of $L_{\bar b}$ by requiring that the {\it quantum gauge} [i.e. anti-BRST] invariant quantities must be independent
of $\theta$ as the Grassmannian variable(s) are {\it only} mathematical artifact and they are {\it not} physical quantity in the 
real sense of the word.

The substitutions of {\it all} the expressions for the secondary variables [cf. Eq. (49)]
into the expansions in (45) lead to the following:
\begin{eqnarray}
&& X_{\mu}^{(ab)} (\tau, \theta) = x_\mu + \theta\,(\bar c\,p_\mu + \bar\beta \,\psi_\mu)\,\equiv x_\mu\,(\tau) 
+ \theta \, (s_{ab} x_\mu),\nonumber\\
&& E^{(ab)} (\tau, \theta) = e(\tau) + \theta \, (\dot {\bar c} + 2\;\bar\beta\; \chi) \equiv e(\tau) + \theta \, (s_{ab} e), \nonumber \\
&& F^{(ab)}(\tau, \theta) = c(\tau) + \theta \, (i\; \bar b) \equiv c(\tau) + \theta \, (s_{ab} c), \nonumber \\
&& {\bar F}^{(ab)}(\tau, \theta) = {\bar c}(\tau) + \theta \, (-\,i\,\bar\beta^2) \equiv {\bar c}(\tau) + \theta \, (s_{ab} \bar c), \nonumber \\
&& \tilde{\beta}^{(ab)}(\tau, \theta) = \bar{\beta}(\tau) + \theta \, (-\,i\,\gamma) \equiv {\beta}(\tau)
+\theta \, (s_{ab} {\beta}), \nonumber \\
&& \Psi_\mu^{(ab)}(\tau, \theta) = {\psi}_\mu(\tau) + \theta \, (i\;\bar\beta\; p_\mu) \equiv {\psi}_\mu(\tau) 
+ \theta \, (s_{ab} \psi_\mu), \nonumber \\
&& \Psi_5^{(ab)}(\tau, \theta) = {\psi}_5(\tau) + \theta \, (i\,\bar\beta\,m) \equiv {\psi}_5(\tau) + \theta \, (s_{ab} \psi_5), \nonumber \\
&& \tilde{\chi}^{(ab)}(\tau, \theta) = {\chi}(\tau) + \bar\theta \, (i\; \dot {\bar\beta}) \equiv {\chi}(\tau) 
+ \theta \, (s_{ab} \chi), \nonumber \\
&&{B}^{(ab)}(\tau, \theta) = {b}(\tau) + \theta \, (2\; i\; \bar\beta\;\gamma) 
\equiv {b}(\tau) + \theta \, (s_{ab} {b}).
\end{eqnarray}
In the above equation, the superscript $(ab)$ on the {\it chiral} supervariables [cf. the l.h.s. of (46) and (50)] denotes the super expansions that have
been derived after the applications of the anti-BRST invariant restrictions in (48). We note that the coefficients of $\theta$, in the above expansions, are 
nothing but the anti-BRST symmetry transformations (12) of our 1D system of a massive spinning relativistic particle.

We wrap up this section with the following comments. First and foremost, we observe that the {\it trivial} anti-BRST invariant
(e.g. $s_{ab}\,p_\mu = s_{ab}\, \gamma = s_{ab} \, {\bar b} = s_{ab}\,\bar\beta = 0 $) variables  have been incorporated in the super
expansions in (46). Second, the non-trivial anti-BRST symmetry transformations (12) have been incorporated in the super expansions (50). Finally,
we have exploited the basic idea of ACSA to BRST formalism where we have demanded that the anti-BRST (i.e. {\it quantum} gauge) invariant quantities
must be {\it independent} of the Grassmannian variable $\theta$ when they are generalized onto the $(1, 1)$-dimensional {\it chiral}
super sub-manifold of the {\it general} (1, 2)-dimensional supermanifold.

\section{Symmetry Invariance of Lagrangians: ACSA}

In this section, we capture the (anti-)BRST symmetry invariance  of the coupled (but equivalent) $L_b$ and $L_{\bar b}$
[cf. Eqs. (20)-(23)] within the framework of ACSA to BRST formalism. In this context, it is pertinent to point out that the 
CF-type condition ($b  + \bar b + 2\,\beta\,\bar\beta = 0$) is responsible for the existence of the coupled (but equivalent)
Lagrangians $L_b$ and $L_{\bar b}$ and {\it it} is {\it also} responsible for the absolute anticommutativity [i.e. $\{s_b, s_{ab}\} = 0$] 
of the (anti-) BRST symmetries ($s_{(a)b}$) {\it and} the absolute anticommutatvity  (i.e. $\{Q_b, Q_{ab}\} = 0$) of the 
corresponding conserved 
$(\dot Q_{(a)b} = 0)$ and off-shell nilpotent (i.e. $\dot Q_{(a)b}^2 = 0)$ (anti-)BRST charges $(Q_{(a)b})$. Thus, it is 
important for us to capture the existence of the (anti-)BRST invariant (i.e. $s_{(a)b}\,[b  + \bar b + 2\,\beta\,\bar\beta] = 0$)  
CF-type restriction in the context of symmetry  considerations of the coupled (but equivalent) Lagrangians  $L_b$ and $L_{\bar b}$
for our 1D system of a reparameterization  invariant {\it massive} model of spinning relativistic particle.

Against the backdrop of the above statements, first of all, we consider the (anti-)BRST symmetry invariance 
[cf. Eqs. (20), (21)] of the Lagrangians  $L_b$ and $L_{\bar b}$. Toward this goal in mind, we generalize 
{\it these} Lagrangians on the (1, 1)-dimensional {\it chiral} and {\it anti-chiral} super sub-manifolds 
(in terms of the corresponding supervariables) as:
\begin{eqnarray}
L_{\bar b} \longrightarrow \tilde L_{\bar b} ^{(c)} (\tau, \theta) & = & \tilde L_{f} ^{(c)} (\tau, \theta) 
+ \bar B^{(ab)} (\tau, \theta)\;\bar B^{(ab)} (\tau, \theta) - \bar B^{(ab)} (\tau, \theta)\,\big[\dot E^{(ab)} (\tau, \theta)\nonumber\\
&-& 2\,\tilde{\bar\beta} ^{(ab)} (\tau, \theta)\,\tilde\beta ^{(ab)} (\tau, \theta)\big] 
- i\,\dot{\bar F}^{(ab)} (\tau, \theta)\,\dot F^{(ab)} (\tau, \theta)\nonumber\\
& + & \tilde{\bar\beta} ^{(ab)} (\tau, \theta)\; \tilde{\bar\beta} ^{(ab)} (\tau, \theta)\; 
\tilde\beta ^{(ab)} (\tau, \theta)\;\tilde\beta ^{(ab)} (\tau, \theta)\nonumber\\
& + & 2\,i\,\tilde{\chi} ^{(ab)} (\tau, \theta)\,\big[\tilde \beta ^{(ab)} (\tau, \theta)\,\dot{\bar F}^{(ab)} (\tau, \theta) 
- \tilde{\bar\beta} ^{(ab)} (\tau, \theta)\,\dot F^{(ab)} (\tau, \theta)\big]\nonumber\\
& + & 2\,E^{(ab)} (\tau, \theta)\big[\dot{\tilde{\bar\beta}} ^{(ab)} (\tau, \theta)\;\tilde{\beta} ^{(ab)} (\tau, \theta)
- \Gamma ^{(ab)} (\tau, \theta)\;\tilde\chi ^{(ab)} (\tau, \theta)\big] \nonumber\\
& + & 2\, \Gamma ^{(ab)} (\tau, \theta) \big[\tilde \beta ^{(ab)} (\tau, \theta)\;{\bar F}^{(ab)} (\tau, \theta) 
-  \tilde{\bar\beta} ^{(ab)} (\tau, \theta)\;{F}^{(ab)} (\tau, \theta)]\nonumber\\
& + & m\,\big[ \tilde{\bar\beta} ^{(ab)} (\tau, \theta)\,  \dot{\tilde{\beta}} ^{(ab)} (\tau, \theta)\,
-  \dot{ \tilde{\bar\beta}} ^{(ab)} (\tau, \theta)\, \tilde{\beta} ^{(ab)} (\tau, \theta)\nonumber\\
& + &  \Gamma^{(ab)} (\tau, \theta)\;\tilde{\chi} ^{(ab)} (\tau, \theta) \big]
 - \dot\Gamma^{(ab)} (\tau, \theta)\;\Psi_5 ^{(ab)} (\tau, \theta),
\end{eqnarray}
\begin{eqnarray}
L_{b} \longrightarrow \tilde L_{b} ^{(ac)} (\tau, \bar\theta) & = & \tilde L_{f} ^{(ac)} (\tau, \bar\theta) 
+ B^{(b)} (\tau, \bar\theta)\;B^{(b)} (\tau, \bar\theta) +  B^{(b)} (\tau, \bar\theta)\,\big[\dot E^{(b)} (\tau, \bar\theta)\nonumber\\
&+& 2\,\tilde{\bar\beta} ^{(b)} (\tau, \bar\theta)\,\tilde\beta ^{(b)} (\tau, \bar\theta)\big] 
- i\,\dot{\bar F}^{(b)} (\tau, \bar\theta)\,\dot F^{(b)} (\tau, \bar\theta)\nonumber\\
& + & \tilde{\bar\beta} ^{(b)} (\tau, \bar\theta)\; \tilde{\bar\beta} ^{(b)} (\tau, \bar\theta)\; \tilde\beta ^{(b)} (\tau, \bar\theta)\;\tilde\beta ^{(b)} (\tau, \bar\theta)\nonumber\\
& + & 2\,i\,\tilde{\chi} ^{(b)} (\tau, \bar\theta)\,\big[\tilde \beta ^{(b)} (\tau, \bar\theta)\,\dot{\bar F}^{(b)} (\tau, \bar\theta) 
- \tilde{\bar\beta} ^{(b)} (\tau, \bar\theta)\,\dot F^{(b)} (\tau, \bar\theta)\big]\nonumber\\
& - & 2\,E^{(b)} (\tau, \bar\theta)\big[{\tilde{\bar\beta}} ^{(b)} (\tau, \bar\theta)\;\dot {\tilde{\beta}} ^{(b)} (\tau, \bar\theta)
+ \Gamma ^{(b)} (\tau, \bar\theta)\;\tilde\chi ^{(b)} (\tau, \bar\theta)\big] \nonumber\\
& + & 2\, \Gamma ^{(b)} (\tau, \bar\theta) \big[\tilde \beta ^{(b)} (\tau, \bar\theta)\;{\bar F}^{(b)} (\tau, \bar\theta) 
-  \tilde{\bar\beta} ^{(b)} (\tau, \bar\theta)\;{F}^{(b)} (\tau, \bar\theta)]\nonumber\\
& + & m\,\big[ \tilde{\bar\beta} ^{(b)} (\tau, \bar\theta)\,  \dot{\tilde{\beta}} ^{(b)} (\tau, \bar\theta)\,
-  \dot{ \tilde{\bar\beta}} ^{(b)} (\tau, \bar\theta)\, \tilde{\beta} ^{(b)} (\tau, \bar\theta)\nonumber\\
& + &  \Gamma^{(b)} (\tau, \bar\theta)\;\tilde{\chi} ^{(b)} (\tau, \bar\theta) \big]
 - \dot\Gamma^{(b)} (\tau, \bar\theta)\;\Psi_5 ^{(b)} (\tau, \bar\theta),
\end{eqnarray}
where the superscript $(c)$ and $(ac)$ on the super Lagrangians (i.e. ${\tilde L}_{\bar b}^{(c)}, {\tilde L}_{b}^{(ac)}$) denote that these
Lagrangians incorporate {\it chiral} and {\it anti-chiral} supervariables that have been obtained {\it after} the (anti-)BRST invariant
restrictions [cf. Eqs. (40), (44), (46), (50)]. Furthermore, we note that the {\it super} first-order Lagrangians are:
\begin{eqnarray}
{\tilde L}_f^{(c)}(\tau, \theta) &=& P_\mu^{(ab)}(\tau, \theta)\,{\dot X}^{\mu{(ab)}}(\tau, \theta) - \frac{1}{2}\,E^{(ab)}(\tau, \theta)\,
\big[P_\mu^{(ab)}(\tau, \theta)\,P^{\mu{(ab)}}(\tau, \theta) - m^2\big] \nonumber \\
& + & \frac{i}{2}\,\big[\Psi_\mu^{(ab)}(\tau, \theta)\,{\dot\Psi}^{\mu{(ab)}}(\tau, \theta) - \Psi_5^{(ab)}(\tau, \theta)\,{\dot\Psi}_5^{(ab)}
(\tau, \theta)\big] \nonumber \\
& + & i\,{\tilde\chi}^{(ab)}(\tau, \theta)\,\big[P_\mu^{(ab)}(\tau, \theta)\,\Psi^{\mu{(ab)}}(\tau, \theta) - m\,\Psi_5^{(ab)}(\tau, \theta)\big],
\end{eqnarray}
\begin{eqnarray}
{\tilde L}_f^{(ac)}(\tau, \bar\theta) &=& P_\mu^{(b)}(\tau, \bar\theta)\,{\dot X}^{\mu{(b)}}(\tau, \bar\theta) - \frac{1}{2}\,E^{(b)}(\tau, \bar\theta)\,
\big[P_\mu^{(b)}(\tau, \bar\theta)\,P^{\mu{(b)}}(\tau, \bar\theta) - m^2\big] \nonumber \\
& + & \frac{i}{2}\,\big[\Psi_\mu^{(b)}(\tau, \bar\theta)\,{\dot\Psi}^{\mu{(b)}}(\tau, \bar\theta) 
- \Psi_5^{(b)}(\tau, \bar\theta)\,{\dot\Psi}_5^{(b)} (\tau, \bar\theta)\big] \nonumber \\
& + & i\,{\tilde\chi}^{(b)}(\tau, \bar\theta)\,\big[P_\mu^{(b)}(\tau, \bar\theta)\,\Psi^{\mu{(b)}}(\tau, \bar\theta) 
- m\,\Psi_5^{(b)}(\tau, \bar\theta)\big].
\end{eqnarray}
It is, at this stage, very essential  to point out that some of the supervariables are (anti-) BRST invariant and,
hence, they are merely {\it ordinary} variables. For instance, we note that: $P_\mu^{(ab)}(\tau, \theta) = 
P_\mu^{(b)}(\tau, \bar\theta) = p_\mu(\tau), \quad \Gamma^{(b)}(\tau, \bar\theta) = \Gamma^{(ab)}(\tau, \theta)
= \gamma(\tau), \quad \tilde{\bar\beta}^{(ab)}(\tau, \theta) = \bar\beta(\tau), \quad {\tilde\beta}^{(b)}(\tau,
\bar\theta) = \beta(\tau),\quad {\bar B}^{(ab)}(\tau, \theta) = {\bar b}(\tau),\quad B^{(b)}(\tau, \bar\theta) = b(\tau)$.

In view of the mappings: $s_b \leftrightarrow \partial_{\bar\theta}, \;\;s_{ab} \leftrightarrow \partial_{\theta}$ [10-12],
we can now capture the (anti-) BRST invariance of $L_{\bar b}$ and $L_b$ [cf. Eqs. (20), (21)] as
\begin{eqnarray}
\frac{\partial}{\partial\,\bar\theta}\,{\tilde L}_b^{(ac)}(\tau, \bar\theta) &=& \frac{d}{d\,\tau}\Big[\frac{c}{2}\,(p^2 + m^2) +
\frac{\beta}{2}\,(p_\mu\,\psi^\mu + m\,\psi_5) +b \,({\dot c} + 2\,\beta\,\chi)\Big] \equiv s_b\,L_b, \nonumber \\
\frac{\partial}{\partial\,\theta}\,{\tilde L}_{\bar b}^{(c)}(\tau, \theta) &=& \frac{d}{d\,\tau}\Big[\frac{\bar c}{2}\,(p^2 + m^2) +
\frac{\bar\beta}{2}\,(p_\mu\,\psi^\mu - m\,\psi_5) - {\bar b} \,({\dot {\bar c}} + 2\,\bar\beta\,\chi)\Big] \equiv s_{ab}\,L_{\bar b}.
\end{eqnarray}
Thus, we have captured the (anti-)BRST invariance of the Lagrangians $L_{\bar b}$ and $L_ b$ [cf. Eq. (21), (20)]
within the framework of ACSA to BRST formalism. Geometrically, the {\it chiral} super Lagrangian
$\tilde L_{\bar b}^{(c)} (\tau,\theta) $ is  a {\it unique} sum of the combination of (super)variables that
have been obtained {\it after} the {\it quantum} gauge (i.e. anti-BRST) invariant restrictions. The translation
of this {\it unique} sum, along the $\theta$-direction of (1, 1)-dimensional {\it chiral} super sub-manifold, 
{\it generates} a total ``time" derivative [cf. Eq. (21)] in the {\it ordinary} space. As a consequence, the 
action integral  $S = \int_{- \infty}^{+ \infty} d\,\tau \, L_{\bar b}$ remains invariant under the anti-BRST
symmetry transformations ($s_{ab}$). In exactly similar fashion, we can discuss the BRST invariance of the Lagrangian
$L_b$ [cf. Eq. (20)]  within the framework of ACSA to BRST formalism and provide the geometrical interpretation for
the super {\it anti-chiral} Lagrangian  $\tilde L_{ b}^{(ac)} (\tau,\bar\theta) $ and its connection with the BRST-invariance 
[cf. Eq. (20)] in the {\it ordinary} space.

Now let us capture Eqs. (22) and (23), where the BRST symmetry transformation operates on $L_{\bar b}$
and the anti-BRST symmetry transformation acts on $L_b$, within the purview of ACSA to BRST formalism.
In this context, let us, first of all, generalize the Lagrangian $L_b$ {\it onto} the {\it chiral} 
(1, 1)-dimensional super sub-manifold such that {\it chiral} supervariables [with the superscript $(ab)$] 
appear in it. In other words, we have the following generalization:
\begin{eqnarray*}
L_{b} \longrightarrow \tilde L_{b} ^{(c)} (\tau, \theta) & = & \tilde L_{f} ^{(c)} (\tau, \theta) 
+ B^{(ab)} (\tau, \theta)\;B^{(ab)} (\tau, \theta) +  B^{(ab)} (\tau, \theta)\,\big[\dot E^{(ab)} (\tau, \theta)\nonumber\\
&+& 2\,\tilde{\bar\beta} ^{(ab)} (\tau, \theta)\,\tilde\beta ^{(ab)} (\tau, \theta)\big] 
- i\,\dot{\bar F}^{(ab)} (\tau, \theta)\,\dot F^{(ab)} (\tau, \theta)\nonumber\\
\end{eqnarray*}
\begin{eqnarray}
& + & \tilde{\bar\beta} ^{(ab)} (\tau, \theta)\; \tilde{\bar\beta} ^{(ab)} (\tau, \theta)
\;\tilde \beta ^{(ab)} (\tau, \theta)\;\tilde\beta ^{(ab)} (\tau, \theta)\nonumber\\
& + & 2\,i\,\tilde{\chi} ^{(ab)} (\tau, \theta)\,\big[\tilde \beta ^{(ab)} (\tau, \theta)\,\dot{\bar F}^{(ab)} (\tau, \theta) 
- \tilde{\bar\beta} ^{(ab)} (\tau, \theta)\,\dot F^{(ab)} (\tau, \theta)\big]\nonumber\\
& - & 2\,E^{(ab)} (\tau, \theta)\big[{\tilde{\bar\beta}} ^{(ab)} (\tau, \theta)\;\dot {\tilde{\beta}} ^{(ab)} (\tau, \theta)
+ \Gamma ^{(ab)} (\tau, \theta)\;\tilde\chi ^{(ab)} (\tau, \theta)\big] \nonumber\\
& + & 2\, \Gamma ^{(ab)} (\tau, \theta) \big[\tilde \beta ^{(ab)} (\tau, \theta)\;{\bar F}^{(ab)} (\tau, \theta) 
-  \tilde{\bar\beta} ^{(ab)} (\tau, \theta)\;{F}^{(ab)} (\tau, \theta)]\nonumber\\
& + & m\,\big[ \tilde{\bar\beta} ^{(ab)} (\tau, \theta)\,  \dot{\tilde{\beta}} ^{(b)} (\tau, \theta)\,
-  \dot{ \tilde{\bar\beta}} ^{(ab)} (\tau, \theta)\, \tilde{\beta} ^{(ab)} (\tau, \theta)\nonumber\\
& + &  \Gamma^{(ab)} (\tau, \theta)\;\tilde{\chi} ^{(ab)} (\tau, \theta) \big]
 - \dot\Gamma^{(ab)} (\tau, \theta)\;\Psi_5 ^{(ab)} (\tau, \theta).
\end{eqnarray}
It should be noted that some of the {\it chiral} supervariables with the superscript $(ab)$ are, primarily, 
the {\it ordinary} variables. For instance, we note that {\it all} the {\it chiral} supervariables on the l.h.s. 
of (46) are {\it actually} such variables [i.e. $\it P_{\mu}^{(ab)} (\tau, \theta) = p_\mu (\tau),
\,\, \Gamma^{(ab)}(\tau,\theta) = \gamma (\tau),\,\, \tilde {\bar \beta}^{(ab)}(\tau,\theta) = \bar \beta (\tau)$].
Keeping in our mind the mapping: $s_{ab} \Leftrightarrow \partial_\theta$, it is clear that we can operate
$\partial_\theta $ on the above {\it chiral} Lagrangian $\tilde L_{b}^{(c)} (\tau, \theta)$ to yield the following:
\begin{eqnarray}
\frac{\partial}{\partial\,\theta}\tilde L_{b}^{(c)} (\tau, \theta)  & = & \frac{d}{d\,\tau}\,\Big[\frac{\bar c}{2}\,(p^2 + m^2)
+ \frac{\bar \beta}{2}\,(p_\mu \, \psi^\mu + m \, \psi_5) + b \, ( \dot{\bar c} + 2\,\bar \beta
\, \chi) + 2\,i\,e\,\bar \beta \gamma\Big] \nonumber\\ 
&-& (\dot{\bar c} + 2\bar \beta \,\chi)\,\Big[\frac{d}{d\,\tau}\,(b+\bar b+2\beta \bar \beta)\Big]\,
+ (2\,i\, \bar \beta \gamma) \;(b+\bar b+2\beta \bar \beta)  \equiv s_{ab}\,L_b.
\end{eqnarray}
The above observation establishes the fact that Lagrangian $L_b$ {\it also} respects the anti-BRST symmetry
transformation (12) provided we invoke the CF-type restriction $(b + \bar b + 2\,\beta\,\bar \beta = 0)$ from 
{\it outside}. In other words, we have captured  the existence of the (anti-)BRST invariant CF-type 
restriction within the framework of ACSA to BRST formalism and have proved that the Lagrangian $L_b$ (which
is {\it perfectly} BRST invariant [cf. Eq. (20)]) is {\it also} invariant w.r.t. the anti-BRST symmetry transformation (12)
provided we confine ourselves to the sub-manifold of the {\it quantum variable} where the CF-type restriction 
is satisfied.

At this juncture, we generalize the {\it ordinary} Lagrangian $L_{\bar b}$ to its counterpart {\it anti-chiral}
Lagrangian $  \tilde L_{\bar b}^{(ac)} (\tau, \bar \theta)$ on the (1, 1)-dimensional {\it anti-chiral} super sub-manifold as:
\begin{eqnarray}
L_{\bar b} \longrightarrow \tilde L_{\bar b} ^{(ac)} (\tau, \bar\theta) & = & \tilde L_{f} ^{(ac)} (\tau, \bar\theta) 
+ \bar B^{(b)} (\tau, \bar\theta)\;\bar B^{(b)} (\tau, \bar\theta) - \bar B^{(b)} (\tau, \bar\theta)\,
\big[\dot E^{(b)} (\tau, \bar\theta)\nonumber\\
&-& 2\,\tilde{\bar\beta} ^{(b)} (\tau, \bar\theta)\,\tilde\beta ^{(b)} (\tau, \bar\theta)\big] 
- i\,\dot{\bar F}^{(b)} (\tau, \bar\theta)\,\dot F^{(b)} (\tau, \bar \theta)\nonumber\\
& + & \tilde{\bar\beta} ^{(b)} (\tau, \bar\theta)\; \tilde{\bar\beta} ^{(b)} (\tau, \bar\theta)\; 
\tilde\beta ^{(b)} (\tau, \bar\theta)\;\tilde\beta ^{(b)} (\tau, \bar \theta)\nonumber\\
& + & 2\,i\,\tilde{\chi} ^{(b)} (\tau, \bar\theta)\,\big[\tilde \beta ^{(b)} (\tau, \bar\theta)\,\dot{\bar F}^{(b)} (\tau, \bar\theta) 
- \tilde{\bar\beta} ^{(b)} (\tau, \bar\theta)\,\dot F^{(b)} (\tau, \bar\theta)\big]\nonumber\\
& + & 2\,E^{(b)} (\tau, \bar\theta)\big[\dot{\tilde{\bar\beta}} ^{(b)} (\tau, \bar\theta)\;\tilde{\beta} ^{(b)} (\tau, \bar\theta)
- \Gamma ^{(b)} (\tau, \bar\theta)\;\tilde\chi ^{(b)} (\tau, \bar\theta)\big] \nonumber\\
& + & 2\, \Gamma ^{(b)} (\tau, \bar\theta) \big[\tilde \beta ^{(b)} (\tau, \bar\theta)\;{\bar F}^{(b)} (\tau, \bar\theta) 
-  \tilde{\bar\beta} ^{(b)} (\tau, \bar\theta)\;{F}^{(b)} (\tau, \bar\theta)]\nonumber\\
& + & m\,\big[ \tilde{\bar\beta} ^{(b)} (\tau, \bar\theta)\,  \dot{\tilde{\beta}} ^{(b)} (\tau, \bar \theta)\,
-  \dot{ \tilde{\bar\beta}} ^{(b)} (\tau, \bar \theta)\, \tilde{\beta} ^{(b)} (\tau, \bar \theta)\nonumber\\
& + &  \Gamma^{(b)} (\tau, \bar\theta)\;\tilde{\chi} ^{(b)} (\tau, \bar\theta) \big]
 - \dot\Gamma^{(b)} (\tau, \bar\theta)\;\Psi_5 ^{(b)} (\tau, \bar\theta).
\end{eqnarray} 
It will be noted that {\it some} of the above anti-chiral supervariables [cf. Eq. (40)] are basically 
{\it ordinary} variables [e.g. $P_\mu ^{(b)} (\tau, \bar\theta) = p_\mu (\tau),\; \tilde\beta ^{(b)} (\tau, \bar\theta)
= \beta (\tau), \; \Gamma ^{(b)} (\tau, \bar\theta) = \gamma (\tau)]$.
As far as the dependence on the Grassmannian variable of $\tilde L_{b}^{(ac)} (\tau, \bar \theta)$
is concerned, it is straightforward to  note that we have $\bar\theta$-dependence. Thus, the mapping:
$s_b \leftrightarrow \partial_{\bar\theta}$ allows us to apply, on the super Lagrangian 
$\tilde L_{\bar b}^{(ac)} (\tau, \bar \theta)$, a derivative ($\partial_{\bar\theta}$) w.r.t. the Grassmannian variable $\bar\theta$. The
ensuing mathematical expression, as the outcome of the above operation,  is as follows:
\begin{eqnarray}
\frac{\partial}{\partial\,\bar\theta}\tilde L_{b}^{(ac)} (\tau, \bar\theta) & = & \frac{d}{d\,\tau}\,\Big[\frac{c}{2}\,(p^2 + m^2)
+ \frac{\beta}{2}\,(p_\mu \, \psi^\mu + m \, \psi_5) - \bar b \, (\dot c + 2\,\beta \, \chi) + 2\,i\,e\,\beta \gamma\Big] \nonumber\\ 
& + & (\dot c + 2\beta \,\chi)\,\Big[\frac{d}{d\,\tau}\,(b+\bar b+2\beta \bar \beta)\Big]\,-(2\,i\, \beta \gamma)\;
(b + \bar b + 2\beta \bar \beta)  \equiv s_{b}\,L_{\bar b}.
\end{eqnarray}
At this stage, it is an elementary exercise to state that the {\it perfectly} anti-BRST invariant   Lagrangian $L_{\bar b}$  [cf. Eq. (21)] also 
respects the BRST symmetry transformations (13) provided the whole theory is considered on 
a sub-manifold of the Hilbert space of quantum variables where the CF-type restriction
$(b + \bar b + 2\,\beta\,\bar\beta = 0)$ is satisfied. In other words, the Lagrangian $L_{\bar b}$ respects the BRST symmetry 
transformations (13) provided we impose the CF-type restriction from {\it outside}. Thus, we have derived  the CF-type 
restriction [cf. Eq. (59)] within the ambit of ACSA to BRST formalism.

We end this section with the remark that the CF-type restriction is the hallmark [5, 6] of a {\it quantum} theory (discussed within the framework 
of BRST formalism). We have shown its existence {\it on} our theory within the framework of ACSA to BRST formalism. 
Hence, we have achieved a proper BRST quantization of our theory of the 1D system.

\section{Off-Shell Nilpotency and Absolute Anticommutativity of the (Anti-)BRST Charges: ACSA}

We have already seen that the (anti-)BRST symmetry transformations (13) and (12) are off-shell nilpotent $(s_{(a)b}^2 = 0)$ and absolutely
anticommuting [cf. Eq. (15)] in nature provided the CF-type restriction [cf. Eq. (14)] is imposed from {\it outside} on our theory.
As the off-shell nilpotent  (anti-)BRST symmetry transformations are generated by the conserved (anti-)BRST charges, 
the above off-shell nilpotency and absolute anticommutativity are {\it also}
respected by the conserved and off-shell nilpotent (anti-)BRST charges. In our present section,
we capture these properties of the conserved ($\dot Q_{(a)b} = 0$) {\it fermionic}  (anti-)BRST charges $Q_{(a)b}$
within the framework of ACSA to BRST formalism.

We have already demonstrated that the Noether conserved charges $Q_{(a)b}^{(1)}$ [cf. Eqs. (26), (27)] are {\it not} off-shell nilpotent.
In fact, they are {\it on-shell} nilpotent. Using the EL-EOMs, we have recast {\it these} conserved charges into another
forms [cf. Eqs. (30), (33)] and denoted them by $Q_{(a)b}^{(2)}$. These latter
forms of the charges turn out to be off-shell nilpotent of order two [i.e. $(Q_{(a)b}^{(2)})^2 = 0$]. 
We now concentrate on $Q_b ^{(2)}$ and  generalize the BRST charge $Q_b^{(2)}$
onto the $(1, 1)$-dimensional {\it anti-chiral} super sub-manifold as:
\begin{eqnarray*}
Q_b^{(2)} \longrightarrow {\tilde Q}_b^{(2)} (\tau, \bar\theta) &=& B^{(b)} (\tau, \bar\theta)\,\dot F ^{(b)} (\tau, \bar\theta) 
- \dot B^{(b)} (\tau, \bar\theta)\,F ^{(b)} (\tau, \bar\theta)\nonumber\\
&+& 2\,i\,E^{(b)} (\tau, \bar\theta)\,\tilde\beta^{(b)} (\tau, \bar\theta)
\,\Gamma^{(b)} (\tau, \bar\theta) +  2\,\tilde\beta^{(b)} (\tau, \bar\theta)\,\tilde{\bar\beta}^{(b)} (\tau, \bar\theta)\,\dot F ^{(b)} (\tau, \bar\theta)\nonumber\\
&-& 2\,F ^{(b)} (\tau, \bar\theta) \,\big[\tilde{\bar\beta}^{(b)} (\tau, \bar\theta)\,\dot{\tilde\beta}^{(b)} (\tau, \bar\theta) + \Gamma^{(b)} (\tau, \bar\theta)\,\tilde\chi ^{(b)} (\tau, \bar\theta)\big]\nonumber\\
\end{eqnarray*}
\begin{eqnarray}
& +& 2\,  B^{(b)} (\tau, \bar\theta)\,\tilde\beta^{(b)} (\tau, \bar\theta)\,
\tilde\chi^{(b)} (\tau, \bar\theta) - 2\,i\,m\,\Gamma^{(b)} (\tau, \bar\theta)\,\tilde\beta^{(b)} (\tau, \bar\theta)\nonumber\\
&-&
\tilde\beta^{(b)} (\tau, \bar\theta)\,\tilde\beta^{(b)} (\tau, \bar\theta)\,\dot{\bar F} ^{(b)} (\tau, \bar\theta)\nonumber\\
&+&  2\,\tilde\chi^{(b)} (\tau, \bar\theta)\,\tilde\beta^{(b)} (\tau, \bar\theta)\,\tilde\beta^{(b)} (\tau, \bar\theta)\,
\tilde{\bar\beta}^{(b)} (\tau, \bar\theta),
\end{eqnarray}
where all the {\it anti-chiral} supervariables with the superscript $(b)$ have been derived in Eqs. (40) and (44). It is evident that there are 
some supervariables in the above expression for ${\tilde Q}_b^{(2)}(\tau, \bar\theta)$ which are {\it actually} ordinary variables [cf. Eq. (40)].
Keeping in our mind the mapping: $s_b \leftrightarrow \partial_{\bar\theta}$ [10-12], it can be explicitly checked that:
\begin{eqnarray}
\frac{\partial}{\partial\,\bar\theta}\,{\tilde Q}_b^{(2)}(\tau, \bar\theta)\; \equiv \; \int d\,\bar\theta
\,\tilde Q_b^{(2)}(\tau, \bar\theta) = 0 \quad \Longleftrightarrow \quad s_b\,Q_b^{(2)} = 0.
\end{eqnarray}
The above relationship is nothing but the proof of the off-shell nilpotency $\big[(Q_b^{(2)})^2 = 0\big]$ of the conserved BRST charge $Q_b^{(2)}$.
To corroborate this statement, we note that, in the {\it ordinary} space, the observation $s_b\,Q_b^{(2)} = 0$ can be mathematically stated as 
Eq. (31) which proves the off-shell nilpotency  $[(Q_b ^{(2)})^2 = 0]$ of the conserved charge $Q_b^{(2)}$.

We now focus on the proof of the nilpotency $\big[(Q_{ab}^{(2)})^2 = 0\big]$ of the anti-BRST charge $Q_{ab}^{(2)}$ [cf. Eq. (33)] within 
the ambit of ACSA to BRST formalism. In this context, we note that $Q_{ab}^{(2)}$ can be generalized onto the $(1, 1)$-dimensional {\it chiral}
super sub-manifold as:
\begin{eqnarray}
Q_{ab}^{(2)} \longrightarrow {\tilde Q}_{ab}^{(2)} (\tau, \theta) &=& \dot{\bar B}^{(ab)} (\tau, \theta)\,\bar F ^{(ab)} (\tau, \theta) 
- \bar B^{(ab)} (\tau, \theta)\,\dot{\bar F} ^{(ab)} (\tau, \theta)\nonumber\\
&+& 2\,i\,E^{(ab)} (\tau, \theta)\,\tilde{\bar\beta}^{(ab)} (\tau, \theta)
\,\Gamma^{(ab)} (\tau, \theta) -  2\,\tilde\beta^{(ab)} (\tau, \theta)\,\tilde{\bar\beta}^{(ab)} (\tau, \theta)\,\dot {\bar F} ^{(ab)} (\tau, \theta)\nonumber\\
&+& 2\,\bar F ^{(ab)} (\tau, \theta) \,\big[\dot{\tilde{\bar\beta}}^{(ab)} (\tau, \theta)\,{\tilde\beta}^{(ab)} (\tau, \theta) - \Gamma^{(ab)} (\tau, \theta)\,\tilde\chi ^{(ab)} (\tau, \theta)\big]\nonumber\\
& - & 2\,  \bar B^{(ab)} (\tau, \theta)\,\tilde{\bar\beta}^{(ab)} (\tau, \theta)\,
\tilde\chi^{(ab)} (\tau, \theta) - 2\,i\,m\,\Gamma^{(ab)} (\tau, \theta)\,\tilde{\bar\beta}^{(ab)} (\tau, \theta)\nonumber\\
&+&
\tilde{\bar\beta}^{(ab)} (\tau, \theta)\,\tilde{\bar\beta}^{(ab)} (\tau, \theta)\,\dot{F} ^{(ab)} (\tau, \theta)\nonumber\\
& - &  2\,\tilde\chi^{(ab)} (\tau, \theta)\,\tilde{\bar\beta}^{(ab)} (\tau,\theta)\,\tilde{\bar\beta}^{(ab)} (\tau, \theta)\,
\tilde{\beta}^{(ab)} (\tau, \theta),
\end{eqnarray}
where {\it all} the supervariables with superscript $(ab)$ are {\it chiral} expansions that have been quoted in Eqs. (46) and (50).
It is pertinent to point out that {\it some} of the {\it chiral} supervariables, on the r.h.s. of (62), are actually {\it ordinary}
variables [cf. Eq. (46)] because they are anti-BRST invariant (e.g. $s_{ab}\,\bar\beta = s_{ab}\,{\bar b} = s_{ab}\,p_\mu = 0$) variables. In
view of the mapping: $s_{ab} \leftrightarrow \partial_\theta$ [10-12], we are in the position to operate a derivative w.r.t. the Grassmannian 
variable $\theta$ on the expression for the {\it super} anti-BRST charge   to show that:
\begin{eqnarray}
\frac{\partial}{\partial\,\theta}\,{\tilde Q}_{ab}^{(2)}(\tau, \theta)\; \equiv \; \int d\,\theta \,\tilde Q_{ab}^{(2)}(\tau, \theta) = 0
\quad \Longleftrightarrow \quad s_{ab}\,Q_{ab}^{(2)} = 0,
\end{eqnarray}
where, as pointed out earlier, we have to substitute the {\it chiral} super expansion (46) and (50) into the r.h.s. of (62) and, then only,
we have to operate $\partial_\theta$. The above equation (63) is nothing but the proof for the off-shell nilpotency $\big[(Q_{ab}^{(2)})^2 
= 0\big]$ of the anti-BRST charge which becomes transparent when we exploit the beauty of the relationship between the continuous symmetries
and their generators as we have shown in Eq. (34). Thus, we have captured the off-shell nilpotency of the anti-BRST charge within the framework
of ACSA.

At this juncture, we pay our attention to capture the absolute anticommutativity of the BRST {\it charge} with the anti-BRST charge using
the theoretical power of ACSA to BRST formalism. In this context, we note that the BRST charge $Q_b^{(2)}$ [cf. Eq. (30)] can be also
generalized {\it onto} the (1, 1)-dimensional {\it chiral} super sub-manifold as
\begin{eqnarray}
Q_b^{(2)} \longrightarrow {\tilde Q}_b^{(2)} (\tau, \theta) &=& B^{(ab)} (\tau, \theta)\,\dot F ^{(ab)} (\tau, \theta) 
- \dot B^{(ab)} (\tau, \theta)\,F ^{(ab)} (\tau, \theta)\nonumber\\
&+& 2\,i\,E^{(ab)} (\tau, \theta)\,\tilde\beta^{(ab)} (\tau, \theta)
\,\Gamma^{(ab)} (\tau, \theta) +  2\,\tilde\beta^{(ab)} (\tau, \theta)\,\tilde{\bar\beta}^{(ab)} (\tau, \theta)\,\dot F ^{(ab)} (\tau, \theta)\nonumber\\
&-& 2\,F ^{(ab)} (\tau, \theta) \,\big[\tilde{\bar\beta}^{(ab)} (\tau, \theta)\,\dot{\tilde\beta}^{(ab)} (\tau, \theta) 
+ \Gamma^{(ab)} (\tau, \bar\theta)\,\tilde\chi ^{(ab)} (\tau, \theta)\big]\nonumber\\
& +& 2\,  B^{(ab)} (\tau, \theta)\,\tilde\beta^{(ab)} (\tau, \theta)\,
\tilde\chi^{(ab)} (\tau, \theta) - 2\,i\,m\,\Gamma^{(ab)} (\tau, \theta)\,\tilde\beta^{(ab)} (\tau, \theta)\nonumber\\
&-&
\tilde\beta^{(ab)} (\tau, \theta)\,\tilde\beta^{(ab)} (\tau, \theta)\,\dot{\bar F} ^{(ab)} (\tau, \theta)\nonumber\\
&+&  2\,\tilde\chi^{(ab)} (\tau, \theta)\,\tilde\beta^{(ab)} (\tau, \theta)\,\tilde\beta^{(ab)} (\tau, \theta)\,
\tilde{\bar\beta}^{(ab)} (\tau, \theta)
\end{eqnarray}
where the {\it chiral} supervariables on the r.h.s. are nothing but the super expansions that have been quoted in (46) and (50). It is 
worthwhile to point out that {\it some} of the supervariables are {\it ordinary} variables in the true sense of the word [cf. Eq. (46)].
We can now operate by the Grassmannian derivative $\partial_\theta$ on (64) to produce:
\begin{eqnarray}
\frac{\partial}{\partial\,\theta}\,{\tilde Q}_b^{(2)}(\tau, \theta) &\equiv& \int d\,\theta\,{\tilde Q}_b^{(2)}(\tau, \bar\theta)
\nonumber \\
& = & i\,(b + {\bar b} + 2\,\beta\,\bar\beta)\,\big[\dot {\bar b} + 2\,\chi\,\gamma 
+ 2\,\beta\,\dot{\bar\beta}\big] - i\,\bar b\,\frac {d}{d\,\tau} \Big[b + {\bar b} + 2\,\beta\,\bar\beta \Big]\nonumber\\
& \equiv & s_{ab} Q_b ^{(2)}  = -\,i\,\{Q_b ^{(2)}, Q_{(ab)}^{(2)}\}. 
\end{eqnarray}
In the above, we have utilized the mapping $\partial_\theta \Leftrightarrow s_{ab}$ to express the l.h.s. in
the {\it ordinary} space. The expression $s_{ab}\,Q_{b}^{(2)}$ is nothing but the absolute anticommutativity (i.e. $\{Q_{b}^{(2)},\,Q_{ab}^{(2)}\}$) of the BRST charge {\it with} the anti-BRST charge $ Q_{ab}^{(2)} $. It is crystal clear that the absolute anticommutativity property (i.e.  $\{Q_{b}^{(2)},\,Q_{ab}^{(2)}\} = 0$) is satisfied if and only if  we impose the condition $(b + \bar b + 2\,\beta\,\bar \beta = 0)$. In other words, we have been able to derive the CF-type restriction: $ b + \bar b + 2\,\beta\,\bar \beta = 0$ (which characterizes a BRST {\it quantized} theory)
by exploiting the theoretical tricks and techniques of ASCA to BRST formalism.

Ultimately,  we concentrate to capture the absolute anticommutativity of the anti-BRST charge $Q_{ab}^{(2)}$ {\it with} the BRST charge $Q_{b}^{(2)}$ within the framework of ACSA to BRST formalism. Toward this goal in mind, we generalize the anti-BRST charge $Q_{ab}^{(2)}$ [cf. Eq. (33)] onto the (1, 1)-dimensional  { \it anti-chiral} super sub-manifold as
\begin{eqnarray}
Q_{ab}^{(2)} \longrightarrow {\tilde Q}_{ab}^{(2)} (\tau, \bar\theta) &=& \dot{\bar B}^{(b)} (\tau, \bar\theta)\,\bar F ^{(b)} (\tau, \bar\theta) 
- \bar B^{(b)} (\tau, \bar\theta)\,\dot{\bar F} ^{(b)} (\tau, \bar\theta)\nonumber\\
& + & 2\,i\,E^{(b)} (\tau, \bar\theta)\,\tilde{\bar\beta}^{(b)} (\tau, \bar\theta)
\,\Gamma^{(b)} (\tau, \bar\theta) -  2\,\tilde\beta^{(b)} (\tau, \bar\theta)\,\tilde{\bar\beta}^{(b)} (\tau, \bar\theta)\,
\dot {\bar F} ^{(b)} (\tau, \bar\theta)\nonumber\\
& + & 2\,\bar F ^{(b)} (\tau, \bar \theta) \,\big[\dot{\tilde{\bar\beta}}^{(b)} (\tau, \bar \theta)\,{\tilde\beta}^{(b)} (\tau, \bar\theta)
- \Gamma^{(b)} (\tau, \bar \theta)\,\tilde\chi ^{(b)} (\tau, \bar\theta)\big]\nonumber\\
& - & 2\,  \bar B^{(b)} (\tau, \bar\theta)\,\tilde{\bar\beta}^{(b)} (\tau, \bar\theta)\,
\tilde\chi^{(b)} (\tau, \bar\theta) - 2\,i\,m\,\Gamma^{(b)} (\tau, \bar \theta)\,\tilde{\bar\beta}^{(b)} (\tau, \bar\theta)\nonumber\\
& + & \tilde{\bar\beta}^{(b)} (\tau, \bar\theta)\,\tilde{\bar\beta}^{(b)} (\tau, \bar\theta)\,\dot{F} ^{(b)} (\tau, \bar\theta)\nonumber\\
& - &  2\,\tilde\chi^{(b)} (\tau, \bar\theta)\,\tilde{\bar\beta}^{(b)} (\tau,\bar\theta)\,\tilde{\bar\beta}^{(b)} (\tau, \bar\theta)\,
\tilde{\beta}^{(b)} (\tau, \bar\theta),
\end{eqnarray}
where the {\it anti-chiral} supervariable on the r.h.s. are nothing but the super expansions (40) and (44) that have been derived after the applications of the {\it quantum} gauge (i.e. BRST) invariant restrictions on the {\it anti-chiral} supervariables. It goes without saying that some of the supervariables on the r.h.s. of (66) are actually {\it ordinary} variables [cf. Eq. (40)]. In view of our understanding that we have: $s_{ab} \Leftrightarrow \partial_{\bar\theta}$ [10-12], we can operate a derivation w.r.t. the Grassmannian variable $\bar \theta$ on the $\tilde Q_{ab}^{(2)} (\tau,\bar \theta)$ as:
\begin{eqnarray}
\frac{\partial}{\partial\,\bar\theta}\,{\tilde Q}_{ab}^{(2)}(\tau, \bar\theta)
&\equiv & \int d\,\bar\theta\,{\tilde Q}_{ab}^{(2)}(\tau, \bar\theta) = 0
\nonumber \\
& = & i\,b\,\frac{d}{d\,\tau}\,\big[b + {\bar b} + 2\,\beta\,\bar\beta\big] - i\,(b + {\bar b} + 2\,\beta\,\bar\beta)\,\big[{\dot b} + 2\,\beta
\,\bar\beta + 2\,\gamma\,\chi \big] \nonumber \\
&\Leftrightarrow &  s_b\,Q_{ab}^{(2)} = -\,i\,\{Q_{ab}^{(2)}, Q_{b}^{(2)}\}.
\end{eqnarray}
In other words [cf. Eq. (38)], we have captured the absolute anticommutativity property of the anti-BRST charge {\it with} the
BRST charge within the framework of ACSA to BRST formalism. A close look at (67) demonstrates that the r.h.s. of
${\partial_{\bar\theta}}\,{\tilde Q}_b^{(2)}(\tau, \bar\theta)$ is {\it zero} if and only if the CF-type restriction is {\it imposed} from
outside. In other words, we have proven the existence of the CF-type restriction: $b + {\bar b} + 2\,\beta\,\bar\beta = 0$ 
(on our BRST {\it quantized} theory) within the framework of ACSA to BRST formalism.

We end this section with the following remarks. First and foremost, we observe that the  mappings:
$s_b \leftrightarrow \partial_{\bar\theta}, \,s_{ab} \leftrightarrow \partial_\theta$ [10-12] play 
crucial role in the proof of the {\it two} decisive  properties of the (anti-)BRST conserved
charges. Second, it is interesting to point out that the conserved charges $Q_{(a)b}^{(2)}$ can
be generalized onto $(1, 1)$-dimensional {\it (anti-)chiral} super sub-manifolds of the {\it general}
$(1, 2)$-dimensional supermanifold. Finally, the operations  of the translational generators
$(\partial_\theta, \partial_{\bar\theta})$ [along the {\it chiral} and {\it anti-chiral} {\it directions}
of the super sub-manifolds] {\it on} the {\it generalized} forms of the super charges ${\tilde Q}_{(a)b}^{(2)}$
lead to the proof of the off-shell nilpotency [cf. Eqs. (61), (63)] and absolute anticommutativity  properties of the (anti-)BRST
charges as well as the deduction  of the (anti-)BRST invariant 
CF-type restriction [cf. Eqs. (65), (67)] within the framework of ACSA to BRST formalism.\\

\section{Conclusions}

In our present endeavour, we have done thread-bare analysis of the {\it classical} gauge, supergauge and reparameterization 
symmetries of the first-order Lagrangian for the 1D system of a {\it massive} spinning relativistic particle. 
We have demonstrated that, in specific limits and identifications, the reparameterization symmetries incorporate the gauge 
and (super)gauge symmetries (cf. Sec. 2). We have established that the constraints of our theory (described by the first-order Lagrangian)
are of first-class variety in the terminology of Dirac's prescription for the classification scheme [36, 37]. We have obtained 
the secondary-constraints from the {\it equivalent} Lagrangians  and corresponding canonical 
Hamiltonians of our 1D system of spinning relativistic particle (cf. Sec. 2 and Appendix A).

We have elevated the combined {\it classical} gauge and supergauge symmetry transformations [cf. Eq. (7)]
to its counterpart {\it quantum} (anti-)BRST symmetry transformations which are respected by the coupled (but equivalent)
(anti-)BRST invariant Lagrangians. The hallmark of a {\it quantum} theory (discussed within the purview of BRST approach)
is the existence of the CF-type restriction(s). We have demonstrated the existence of a {\it single} CF-type restriction {\it on} 
our theory by demanding the absolute anticommutativity of the off-shell nilpotent (anti-)BRST symmetry transformations 
as well as by proving the {\it equivalence} of the coupled Lagrangians with respect to the {\it quantum}
gauge [i.e. (anti-)BRST] symmetry  transformations [cf. Eqs. (22), (23)]. In other words, the absolute
anticommutativity of the (anti-)BRST symmetry transformations and existence of the coupled (but equivalent)
Lagrangians {\it owe} their {\it origins} to the CF-type restriction which defines a sub-manifold in
the quantum Hilbert space of variables that is defined by the equation: $b + {\bar b} + 2\,\beta\,\bar\beta = 0$.

To corroborate the sanctity of our (anti-)BRST symmetry transformations, coupled (but equivalent)
Lagrangians and their invariance(s), we have exploited the theoretical potential and power of ACSA
to BRST formalism [20-24] where {\it only} the (anti-)chiral supervariables and their suitable expansion(s)
along the Grassmannian direction(s) have been considered. One of the novel observations, in this context, 
has been the proof of absolute anticommutativity of the conserved and nilpotent (anti-)BRST charges 
within the framework of ACSA to BRST formalism where  we have considered {\it only} the (anti-)chiral super 
expansions. This proof, it should be emphasized, is {\it obvious} when the {\it full} expansions of the supervariables (defined on the 
{\it full} supermanifold) are taken into account. The importance of the ACSA to BRST formalism lies in its
simplicity and its dependence on the quantum gauge [i.e. (anti-)BRST] invariant restrictions on the 
supervariables which are defined on the (anti-) chiral super sub-manifolds of the {\it general} (full) supermanifold.
Whereas the (anti-)chiral super sub-manifolds are characterized by a {\it single} Grassmannian variable, 
the {\it general} supermanifold is defined by the superspace coordinates that incorporate a {\it pair} of Grassmannian 
variables. The {\it quantum} (anti-)BRST symmetry transformations are found to be associated with the translational  
generators $(\partial_\theta, \partial_{\bar\theta})$ along the Grassmannian directions $(\theta, \bar\theta)$.

We have devoted  a great of discussion on the derivation and proof of the existence of CF-type restriction (see, also, e.g. [25]) on our theory
because the hallmark [5,6] of a {\it quantum} theory (described and discussed within the framework of BRST formalism) is {\it its}
presence. We have shown its appearance in the context of absolute anticommutativity of the (anti-)BRST symmetries [cf. Eq. (14)],
invariance and equivalence of the coupled (but equivalent) Lagrangians [cf. Eqs. (23)-(25)] and absolute anticommutativity of
the conserved and nilpotent (anti-)BRST charges [cf. Eqs. (35), (37)] in the {\it ordinary} space. These features have also
been captured in the {\it superspace} by exploiting the theoretical potential and power of ACSA to BRST formalism (cf. Secs. 6, 7).

We would like to comment on the various kinds of superfield approaches (e.g. USFA, AVSA, ACSA) to BRST formalism that have been 
developed over the years. The USFA is the one where mathematically beautiful HC has been exploited to derive the 
(anti-) BRST symmetry transformations for the gauge and associated (anti-)ghost fields in the case of a (non-)Abelian 1-form gauge theory (see. e.g. [10-12]).
In addition, it has led to the systematic derivation of the (anti-)BRST invariant CF-condition [7]. The AVSA is a minor extension of the
USFA (developed in [10-12]) where the HC and gauge invariant restriction(s) play an important role {\it together} for the derivation of the
proper (anti-)BRST symmetry transformations of the gauge, (anti-)ghost and matter fields {\it together} for an interacting gauge theory. The 
ACSA is a simplified version of USFA where the {\it quantum} gauge [i.e. (anti-)BRST] invariant restrictions on the 
superfields/supervariables lead to the derivation of
the (anti-)BRST symmetry transformations for {\it all} the fields. Within the framework of ACSA, the (anti-)BRST invariant CF-type restriction(s)
arise in the proof of (i) the invariance of the coupled (but equivalent) Lagrangian densities, and (ii) the absolute anticommutstivity of the
conserved and nilpotent (anti-)BRST charges.

In our present investigation, we have performed the BRST and supervariable analysis of a toy model 
(i.e. 1D system) of a {\it massive} spinning relativistic particle where {\it only} the (super)gauge symmetry transfromations (7)
have been eploited for the BRST analysis. We have {\it not} devoted any time on the BRST analysis
corresponding to the infinitesimal raparameterization symmetries (4). It would be a nice future endeavor to exploit the
{\it latter} symmetry transformations for the BRST analysis in view of the fact that {\it such} an exercise has already been performed by us in the case of a
scalar relativistic particle [26].   We plan to extend the richness 
of our theoretical analysis to the  realm of interesting  systems of quantum field theory as well as 
diffeomorphism invariant theories of gravitation and (super)strings. It is worthwhile to mention here that, in a recent set of papers (see, e.g. [38-40]),
the BRST  analysis has been performed for the celebrated ABJM theory. It would be, therefore, a very nice future project for us to apply
our present theoretical analysis to the ABJM theory.
We are currently very seriously involved with
the {\it classical} diffeomorphism symmetry and its elevation to the {\it quantum} (anti-)BRST 
symmetries for the system of scalars, vectors and metric tensor. Our results, in this direction,
would be reported elsewhere [41].\\

\begin{center}
{\bf Appendix A: On the Derivation of Secondary Constraints}\\
\end{center}
\vskip 0.5cm
The purpose of our present Appendix is to derive the secondary constraints $p^2 - m^2 \approx 0$ and 
$p_\mu\, \psi^\mu - m\,\psi_5\approx 0$ from all the  {\it three} equivalent Lagrangians (1) as well as from the corresponding Hamiltonians 
(11). First of all, we focus on $L_0$ and $H_c ^{(0)}$. It is evident that the expression for the canonical conjugate 
momenta $(p_\mu)$ w.r.t. the coordinate $(x^\mu)$ is:
\[p_\mu  = \frac {\partial L_0}{\partial \dot x^\mu} = \frac {m\,(\dot x_\mu + i\,\chi\,\psi_\mu)}
{\sqrt{(\dot x^\rho + i\,\chi\,\psi^\rho)\,(\dot x_\rho + i\,\chi\,\psi_\rho)}}.\eqno (A.1)\]
It is self-evident that the Euler-Lagrange equation of motion (EL-EOM) for our free system is $\dot p_\mu  = \frac {d p_\mu}{d\tau} = 0$
and it satisfies the mass-shell condition:
\[p_\mu \,p^\mu  = \frac {m^2\,(\dot x_\mu + i\,\chi\,\psi_\mu)\,(\dot x^\mu + i\,\chi\,\psi^\mu)}
{\big(\sqrt{(\dot x^\rho + i\,\chi\,\psi^\rho)\,(\dot x_\rho + i\,\chi\,\psi_\rho)} \big)^2} = m^2. \eqno (A.2)\]
Furthermore, it is evident from $L_0$ that we have the expression for the 
canonical conjugate momentum w.r.t. the variable $\chi$ (with $\Pi_\chi$ weakly equal to zero) as:
\[\Pi_\chi = \frac{\partial L_0}{\partial {\dot \chi}} = 0 \qquad \implies \qquad \Pi_\chi \approx 0.\eqno (A.3)\]
Thus, we have the primary constraint as $\Pi_\chi \approx 0$ (weekly zero). Hence, we are allowed to take a first-order time derivative on it in the
EL-EOM w.r.t. $\chi$ as (see, e.g. [36, 37, 42]):
\[
\frac{d}{d\,\tau}\Big(\frac{\partial L_0}{\partial {\dot \chi}}\Big) = \frac{\partial L_0}{\partial {\chi}} \qquad \implies \qquad 
{\dot\Pi}_\chi = +\, i \,  (p_{\mu}\, \psi^{\mu} - m \, \psi_5)  \approx 0,\eqno (A.4)
\]
which leads to the derivation of the secondary constraint as $(p_{\mu}\, \psi^{\mu} - m \, \psi_5)  \approx 0$. The same result is also obtained
from the canonical Hamiltonian $H_c^{(0)}$ as we note that the Heisenberg EOMs for the time derivative on the 
conjugate momenta operators ($p_\mu, \Pi_\chi $) are:
\[
{\dot p}_\mu = - \, i \, \big[p_\mu, H_c^{(0)}\big] = 0,\qquad 
{\dot \Pi}_\chi = - \, i \, \big[\Pi_\chi, H_c^{(0)}\big] =  i \,  (p_{\mu}\, \psi^{\mu} - m \, \psi_5)  \approx 0.\eqno (A.5)
\]
Hence, we have derived the secondary constraints $(p^2 - m^2) \approx 0$ and $(p_{\mu}\, \psi^{\mu} - m \, \psi_5)  \approx 0$ from the
Lagrangian $L_0$ with square-root and the corresponding canonical Hamiltonian $H_c^{(0)}$ (which, primarily,  is nothing
but the secondary constraint  on our theory).

Now the stage is set to derive the primary and secondary constraints from the Lagrangians $L_f$ and $L_s$ [cf. Eq. (1)] and corresponding 
canonical Hamiltonian $H_c$ [cf. Eq. (11)]. It is evident that the expressions for the canonical conjugate momenta  w.r.t. the variables $e$
and $\chi$, from $L_f$ and $L_s$, are
\[\Pi_e = \frac{\partial L_r}{\partial \, {\dot e}} \approx 0, \qquad \quad  \Pi_\chi = \frac{\partial L_r}{\partial
\, {\dot \chi}} \approx 0, \qquad \quad r = f, s.   \eqno (A.6)\]
Hence, we have primary constraints $\Pi_e \approx 0$ and $\Pi_\chi \approx 0$ (i.e. weekly zero).
As per the Dirac prescription (see, e.g. [36, 37, 42]), we are allowed to take a
first-order time derivative on these primary constraints. Using the EL-EOMs w.r.t. $e$ and $\chi$ variables, we find that:
\[
\frac{d}{d\,\tau}\Big(\frac{\partial L_f}{\partial\, {\dot e}}\Big) = \frac{\partial L_f}{\partial\,e} \qquad \implies \qquad 
{\dot \Pi}_e = -\, \frac{1}{2}\, (p^2 - m^2) \approx 0,
\]
\[~~~~~~~~~~~~~\frac{d}{d\,\tau}\Big(\frac{\partial L_f}{\partial\, {\dot \chi}}\Big) = \frac{\partial L_f}{\partial\,\chi}\qquad \implies
\qquad 
{\dot \Pi}_\chi = -\,i \, (p_{\mu}\, \psi^{\mu} - m \, \psi_5) \approx 0.\eqno (A.7)
\]
It is clear, from the above, that we have already derived the secondary constraints $ p^2-m^2\approx\,0$ and $ p_\mu\, \psi^\mu - m\,\psi_5\approx 0$. As far as the Lagrangian $L_s$ is concerned, we have the expression for the canonical conjugate momenta  $(p_\mu)$ w.r.t. the coordinates $(x^\mu)$ as:
\[
p_\mu = \frac{\partial\, L_s}{\partial\, \dot{x^\mu}} = \frac{(\dot x_\mu + i\,e\,\psi_\mu)}{e}. \eqno (A.8)
\]
 The EL-EOM w.r.t. $e$ (from the second-order Lagrangian $L_s$) yields
\[
\frac{d}{d\tau}\Big(\frac{\partial L_s}{\partial \dot e}\Big)\,=\,\frac{\partial L_s}{\partial e} \qquad \Longrightarrow \qquad \dot{\Pi_e} =  -\frac{1}{2}\,\frac{(\dot x_\mu+i\,\chi\,\psi_\mu)(\dot x^\mu+i\,\chi\,\psi^\mu)}{e^2}\,+ \, \frac{m^2}{2}\]
\[~~~~~~~~~~~~~~~~~~~~~\equiv  -\,\frac{1}{2}(p^2 - m^2)\approx 0,
\eqno (A.9)\]
which produces   the secondary  constraint $p^2 - m^2\,\approx \,0$. Similarly, the EL-EOM w.r.t. $\chi$ is
\[
\frac{d}{d\tau}\Big(\frac{\partial L_s}{\partial \dot \chi}\Big) = \frac{\partial L_s}{\partial \chi}
\qquad \Longrightarrow \qquad \dot{\Pi_{\chi}} = -i\, (p_\mu\, \psi^\mu - m\,\psi_5) \approx 0.
\eqno (A.10)\]
Hence, we have derived ${\it both}$ the secondary constraints $p^2 - m^2\,\approx \,0$ and 
$ p_\mu\, \psi^\mu - m\,\psi_5\approx 0$ from the first- and second-order Lagrangians $L_f$ and $L_s$, respectively.

Against the backdrop of the existence of the primary constraints $\Pi_e\approx0$ and $\Pi_{\chi}\approx 0$, we derive the secondary constraints from the canonical Hamiltonian $ {H_c}$ [ cf. Eq. (11)] as follows (with the natural units $\hbar = c = 1$), namely;
\[
\dot \Pi_e\,=\,-\,i\,\big[\Pi_e,\,H_c\big] = - \,\frac{1}{2}\,(p^2 - m^2)\approx 0,\]
\[~~~~~~~\dot \Pi_\chi\,=\,- \, i \, \big[\Pi_\chi,\, H_c\big] =  -\,i \,  (p_{\mu}\, \psi^{\mu} - m \, \psi_5)  \approx 0,
\eqno (A.11)\]
where we have used the canonical commutator $[e,\,\Pi_e]= i$ and canonical anticommutator as:
$\{\chi,\, \Pi_\chi \}= i$ in the natural units $\hbar = c = 1$. We end this Appendix with the 
remark that we have derived the secondary constraints $p^2 - m^2\approx 0$ and $ p_\mu\, \psi^\mu 
- m\,\psi_5\approx 0$ from all the {\it three} equivalent Lagrangian (1) as well as from the 
canonical Hamiltonian (11). As a passing remark, we note that the whole dynamics of our theory
is governed by the secondary constrains because a close look at $H_c$ [cf. Eq. (11)] demonstrates
that the Hamiltonian is a {\it linear} combination of the constraints  $p^2 - m^2\approx 0$ and $ p_\mu\, \psi^\mu - m\,\psi_5\approx 0$.
Last but not least, we note that the constraint $p_\mu\,\psi^\mu - m\, \psi_5 \approx 0$ is the square-root of the mass-shell
condition $p^2 - m^2 \approx 0$ because we observe the following:
\[(p_\mu\,\psi^\mu - m\, \psi_5)^2 = \frac {1}{2}\,\{p_\mu\,\psi^\mu - m\, \psi_5, \; p_\nu\,\psi^\nu - m\, \psi_5\}. \eqno (A.12)\]
It is straightforward to note, from the first-order and second-order Lagrangians, that we have the following explicit expressions, namely;
\[\Pi^\mu_{(\psi)} = -\, \frac{i}{2}\,\psi^\mu, \qquad\qquad \Pi_{(\psi_5)} = \frac{i}{2}\,\psi_5, \eqno (A.13)\]
as the canonical conjugate momenta w.r.t. the fermionic variables $\psi_\mu$ and $\psi_5$. As a consequence, we 
have the following canonical anticommutators:
\[\{ \psi_\mu, \psi_\nu\} = -\,2\,\eta_{\mu\nu}, \qquad\qquad \{ \psi_5, \psi_5\} = 2 \; \Longrightarrow \;\psi_5 ^2 = 1. \eqno (A.14)\]
Using the above anticommutators (A.14), we find that the r.h.s. of (A.12) is nothing but the mass-shell condition: $p^2 - m^2 = 0.$
This observation establishes the fact that the {\it two} secondary constraints (i.e. $p^2- m^2 \approx 0,
\; p_\mu\,\psi^\mu - m\,\psi_5 \approx 0$)  of the theory are inter-related.\\

\begin{center}
{\bf Appendix B: On the Derivation of Conserved Noether Charges}\\
\end{center}
\vskip 0.4cm
The central goal of our present Appendix is to derive the (anti-)BRST charges $Q_{(a)b}^{(1)}$ from the
{\it basic } principle of  Noether's theorem and prove {\it their} conservation law by exploiting
the EL-EOMs that emerge out from the coupled (but equivalent) Lagrangians $L_b$ and $L_{\bar b}$. First
of all, we concentrate on the derivation of the anti-BRST charge $Q_{ab}^{(1)}$ by using the following
standard formula for the Noether charge in 1D, namely;
\[ Q_{ab}^{(1)} = (s_{ab}  x_\mu) \Big (\frac{\partial L_{\bar b}}{\partial \dot x_\mu}\Big) + 
(s_{ab} \psi_\mu) \Big (\frac{\partial L_{\bar b}}{\partial \dot \psi_\mu}\Big) + (s_{ab} \psi_5) \Big (\frac{\partial L_{\bar b}}{\partial \dot \psi_5}\Big) + (s_{ab} e) \Big (\frac{\partial L_{\bar b}}{\partial \dot e}\Big) + (s_{ab}\beta) \Big (\frac{\partial L_{\bar b}}{\partial \dot \beta}\Big)\]
\[ + (s_{ab} c) \Big (\frac{\partial L_{\bar b}}{\partial \dot c}\Big) +  (s_{ab}\bar c) \Big (\frac{\partial L_{\bar b}}{\partial \dot{\bar c}}\Big) - \frac {\bar c}{2} (p^2 + m^2) -\frac{\bar\beta}{2}\,(p_\mu\,\psi^\mu + m\,\psi_5) +\bar b \, (\dot{\bar c} + 2\,\bar\beta\,\chi),
\eqno (B.1)\]
where we have taken into account the transformation property of the Lagrangian [cf. Eq. (21)] under the anti-BRST
symmetry transformation $(s_{ab})$ that is quoted in Eq. (12). Furthermore, we have utilized our knowledge of 
the transformations: $s_{ab}\bar\beta = 0$ and $s_{ab}\gamma = 0$. In addition, we have also noted that  
$\Big ({\partial L_{\bar b}}/{\partial \dot \chi}\Big) = 0$ and $\Big ({\partial L_{\bar b}}/{\partial \dot b}\Big) = 0$. 
We would like to lay emphasis on the fact that we have taken into account the convention of the left-derivative
w.r.t. the fermionic  variables ($\psi_\mu, \psi_5, c, \bar c$). Hence, the expression for the anti-BRST charge in Eq. (B.1)
is correct according to our adopted convention of the derivatives.

The substitutions of the transformations (12) and the proper expressions for the derivatives  into (B.1)
lead to the exact expression for the anti-BRST charge  ($Q_{ab}^{(1)})$ that has been quoted in Eq. (26).
The conservation law (i.e $\dot Q_{ab}^{(1)}=0)$ can be proven by taking into account the following EL-EOMs
that emerge out from the Lagrangian $L_{\bar b}$, namely;
\[
\dot p_{\mu}\,=\,0, \qquad \dot{x_\mu}\,=\, e\,p_\mu - i\,\chi\,\psi_\mu,\qquad\dot\psi_\mu = \chi\,p_\mu, \quad
2\,\bar b - \dot e + 2\,\beta\,\bar\beta = 0, \quad \]
\[\dot\psi_5 = m\,\chi -i\,\dot\gamma \; \equiv \; 2\,e\,\chi- m\,\chi -2\,(\beta\,\bar c- \bar\beta\,c), \quad
\ddot c + 2\,\beta\,(\dot\chi + i\,\gamma) + 2\,\dot\beta\,\chi = 0,\]
\[\ddot{\bar c} + 2\,\bar\beta\, (\dot\chi + i\,\gamma) + 2\,\chi\,\dot{\bar\beta}  = 0,\quad
(m - e)\,\dot{\bar\beta} = \bar\beta\,(\bar b + \beta\,\bar\beta) + i\,\chi\,\dot{\bar c} + \gamma\,\bar c,\]
\[(e - m)\,\dot\beta = \beta\,(\bar b + \beta\,\bar\beta - \dot e) - i\,\chi\,\dot c + \gamma\,c.
\eqno (B.2)\]
It would be noted that we have not incorporated, in the above, the EL-EOMs w.r.t. the
variables $e$ and $\chi$ because these have already been invoked and utilized in Eq. (32).

At this stage, we focus on the derivation of the Noether  conserved charge $Q_{b}^{(1)}$ [cf. Eq. (27)].
Applying the basic concept behind the Noether theorem, we note that we have the following expression for the BRST charge, namely;
\[ Q_{b}^{(1)} = (s_{b}  x_\mu) \Big (\frac{\partial L_{ b}}{\partial \dot x_\mu}\Big) + 
(s_{b} \psi_\mu) \Big (\frac{\partial L_{b}}{\partial \dot \psi_\mu}\Big) + (s_{b} \psi_5) \Big (\frac{\partial L_{ b}}{\partial \dot \psi_5}\Big) 
+ (s_{b} e) \Big (\frac{\partial L_{b}}{\partial \dot e}\Big) + (s_{b}\bar\beta) \Big (\frac{\partial L_{ b}}{\partial \dot {\bar\beta}}\Big)\]
\[ + (s_{b} c) \Big (\frac{\partial L_{ b}}{\partial \dot c}\Big) +  (s_{b}\bar c) \Big (\frac{\partial L_{ b}}{\partial \dot{\bar c}}\Big) - \frac {c}{2} \,(p^2 + m^2) -\frac{\beta}{2}\,(p_\mu\,\psi^\mu + m\,\psi_5) - b \, (\dot{c} + 2\,\beta\,\chi),
\eqno (B.3)\] 
where we have taken into account the {\it trivial} transformations $s_b \, \beta = 0, s_b \, \gamma = 0$ as well as the fact that
$({\partial \, L_b}/{\partial \, {\dot \chi}}) = 0$ and $({\partial \, L_b}/{\partial \, {\dot{\bar b}}}) = 0$. The substitutions of
the transformations (13) and the proper and precise values of the derivatives (derived from $L_b$) lead to the derivation of the 
conserved charge $Q_b^{(1)}$ that has been listed in Eq. (27). The conservation law ${\dot Q}_b^{(1)} = 0$ can be proven by taking into
account the following EL-EOMs that emerge out from the Lagrangian $L_{b}$, namely;
\[{\dot p}_\mu = 0, \qquad {\dot x}_\mu = e \, p_\mu - i \, \chi \, \psi_\mu, \qquad {\dot \psi}_\mu = \chi \, p_\mu, \qquad 2\,b + {\dot e} + 2\,\beta\,\bar\beta = 0,  \]
\[\dot\psi_5 = 2 \, e \, \chi - 2\,(\beta \, {\bar c} - \bar\beta \, c) - m \, \chi \equiv m \, \chi - i \, \dot\gamma, \quad \ddot {\bar c} + 2\,\bar\beta\, (\dot\chi\,+ i\,\gamma) + 2\,\chi\,\dot{\bar\beta} = 0,\] 
\[
\ddot{c} + 2\,\beta\, (\dot \chi - i\,\gamma) + 2\,\dot\beta\,\chi, \qquad (m - e)\,{\dot\beta} + b\,\beta + \bar\beta\,{\beta}^2 - i \, \chi\,{\dot c} - \gamma\,c = 0, \]
\[(m - e)\,\dot{\bar\beta} - (b + {\dot e})\,\bar\beta - {\bar\beta}^2\,{\beta} - i \, \chi\,\dot{\bar c} - \gamma\,{\bar c} = 0.
 \eqno (B.4)\]
We would like to point out that, in the above, we have {\it not} incorporated 
a couple of equations [hidden in Eq. (29)] which are the EL-EOMs w.r.t. the variables 
$e$ and $\chi$ from $L_b$.

We end this Appendix with the remarks that we have derived the Noether conserved (anti-)BRST 
charges $Q_{(a)b}^{(1)}$ [cf. Eqs. (26), (27)] which are
{\it only} {\it on-shell} nilpotent of order two. To accomplish the off-shell nilpotency
(without any use of EL-EOMs and/or CF-type restriction), it is essential
for us to use the EL-EOMs w.r.t. $e$ and $\chi$ [cf. Eqs. (29), (32)] to convert the Noether conserved (anti-)BRST charges
$(Q_{(a)b}^{(1)})$ into $Q_{(a)b}^{(2)}$ [cf. Eqs. (30), (33)].\\

\begin{center}
{\bf Appendix C: On the Step-by-Step Derivation of the BRST Symmetry Transformations by Using the ACSA to BRST Formalism}\\
\end{center}
\vskip 0.5cm
In our present Appendix, we derive the BRST symmetry transformations (13) by exploiting the theoretical potential and power of ACSA to BRST
formalism in a systematic fashion. In other words, we obtain the relationships in (43) from the BRST invariant restrictions on the {\it anti-chiral} supervariables in (42). In this context, we note that:
\[
\tilde{\bar\beta}(\tau, \bar\theta) \, \Gamma^{(b)}(\tau, \bar\theta) = \bar\beta(\tau) \,\gamma(\tau) \;\; \implies \;\; [\bar\beta(\tau) + 
\bar\theta \, f_3 (\tau)]\,\gamma(\tau) = \bar\beta(\tau)\,\gamma(\tau),    \eqno{(C.1)} \]
implies that $f_3 (\tau)\propto \gamma(\tau)$. In other words, we obtain: $f_3(\tau) = k\;\gamma(\tau)$ where $k$ is a numerical constant. Now we 
focus on the following restriction on the combination of the {\it anti-chiral} supervariables, namely;
\[
B^{(b)}(\tau, \bar\theta)\,\tilde{\bar\beta}(\tau, \bar\theta) + \Gamma^{(b)}(\tau, \bar\theta)\,{\bar F}(\tau, \bar\theta) = b(\tau)\,
\bar\beta(\tau) + \gamma(\tau)\,{\bar c}(\tau),   \eqno{(C.2)}
\]
which is precisely the generalization of the BRST invariant quantity: $s_b(b\,\bar\beta + \gamma \, {\bar c}) = 0$. The substitution of the expansion
for $\tilde{\bar\beta}(\tau, \bar\theta) = \bar\beta(\tau) + \bar\theta \, [k\,\gamma(\tau)]$ into the above and use of Eq. (40) yield the 
following relationship:
\[
b(\tau) \, \big[\bar\beta(\tau) + \bar\theta\,(k\,\gamma)\big] + \gamma(\tau)\,\big[{\bar c}(\tau) + \bar\theta \, (b_2(\tau))\big] = b(\tau)\,\bar\beta(\tau)
+ \gamma(\tau)\,{\bar c}(\tau).     \eqno{(C.3)}
\]
It is clear, from the above, that we obtain the following:
\[
k\,b(\tau)\,\gamma(\tau) - \gamma\,b_2(\tau) = 0 \implies \big[k\,b(\tau) - b_2(\tau)\big]\,\gamma(\tau) = 0.   \eqno{(C.4)}
\]
From this relationship, it is clear that if we wish to have $s_b \, {\bar c} = i\,b$ in our theory, the constant $k$ would turn out to be: $k = i$.
It would be recalled that $s_b \, {\bar c} = i\,b$ is a standard transformation w.r.t. the BRST symmetry $(s_b)$  within the framework of BRST formalism. Thus, ultimately, we have derived the following super expansions in terms of (13), namely;
\[
{\bar F}^{(b)} (\tau, \bar\theta) = {\bar c}(\tau) + \bar\theta\,(i\,b(\tau)) \equiv {\bar c}(\tau) + \bar\theta\,(s_b\,{\bar c}(\tau)),
\]
\[
\tilde{\bar\beta}^{(b)} (\tau, \bar\theta) = {\bar\beta}(\tau) + \bar\theta\,(i\,\gamma (\tau)) \equiv {\bar\beta}(\tau) + \bar\theta\,
(s_b\,{\bar\beta}(\tau)).    \eqno{(C.5)}
\]
Thus, we note that coefficients of $\bar\theta$ are nothing but the BRST symmetry transformations: $s_b c  =i\,b,\; s_b \bar\beta = i\,\gamma$.
Now we concentrate on the BRST invariant quantity: $s_b\,(\beta ^2 \, \bar \beta + c\,\gamma)=0.$ 
This can be generalized onto the (1, 1)-dimensional {\it anti-chiral} super sub-manifold as:
\[
\tilde\beta^{2\,(b)}(\tau,\bar \theta)\,\tilde{\bar\beta}^{(b)}(\tau,\bar \theta) + F(\tau,\bar \theta)\,\Gamma^b (\tau,\bar \theta) \equiv \beta ^2 (\tau)\, \bar \beta (\tau) + c (\tau) \, \gamma (\tau).
\eqno (C.6)
\]
The substitutions  of $\Tilde\beta^{(b)} (\tau,\bar \theta) = \beta (\tau),\,\, \Gamma^b (\tau,\bar \theta) = \gamma (\tau), \;  F(\tau, \bar\theta) = c(\tau) + \bar \theta\,b_1(\tau)$  and (C.5) lead to the following explicit expressions for the l.h.s. and r.h.s., namely;
\[
{\beta}^2 (\tau)\big[{\bar \beta (\tau) + \bar \theta (i\,\gamma (\tau))}  \big] + [c(\tau) + \bar \theta \, b_1(\tau)]\,\gamma (\tau) = {\beta}^2 (\tau)\,{\bar \beta} (\tau) + c(\tau)\,\gamma (\tau).
\eqno (C.7)
\]
The straightforward algebra yields: $ b_1 (\tau) = -\,i\,\beta^2 (\tau).$ Hence, we have the following super expansion
for the {\it anti-chiral} supervariable $F(\tau, \bar\theta)$, namely;
\[
F^{(b)} (\tau,\bar \theta) = c(\tau) + \bar \theta\, (-\,i\,\beta^2 (\tau)) = c(\tau) + \bar\theta\,(s_b \, c(\tau)).
\eqno (C.8)
\]
We note that the coefficient of $\bar\theta$ is nothing but $s_b c = -\,i\,\beta ^2$.
It is now straightforward to note that, from the BRST invariant quantities: $ s_b \,(\dot c 
+ 2\,\beta \, \chi) = 0$ and $ s_b\,( \bar b + 2\, \beta \bar \beta) = 0$ and {\it their} 
generalizations onto the (1, 1)-dimensional {\it anti-chiral} super sub-manifold [cf. Eq. (42)]
(with the {\it inputs} from (C.5) and (C.8)), lead to the following:
\[
f_1 (\tau) = \dot c + 2\, \beta\,\chi, \,\qquad f_5 = -\,2\,i\,\beta\,\gamma.
\eqno (C.9)
\]
The above values of the secondary variables yield the following super expansions:
\[
E^{(b)} (\tau, \bar \theta) = e(\tau) + \bar \theta \,[ \dot c (\tau) + 2\,\beta(\tau)\,\chi (\tau)] \equiv e(\tau) + \bar\theta \,(s_b\, e(\tau)),
\]
\[\bar B^{(b)} (\tau, \bar \theta) \equiv \bar b (\tau) + \bar \theta\, (-\,2\,i\,\beta(\tau)\,\gamma (\tau)) \equiv \bar b (\tau) + \bar \theta \,(s_b \bar b (\tau)).
\eqno (C.10)
\]
Hence, we have derived the BRST symmetry transformations: $ s_b\,e = \dot c + 2\,\beta \, \chi,\,\, s_b \, \bar b 
= -\,2\,i\,\beta\,  \gamma $ as the coefficients of $\bar\theta$. Now the {\it inputs } from
Eqs. (C.5), (C.8) and (C.10) lead to the determination of {\it all} the {\it secondary variables} 
[cf. Eq. (43)] from the BRST invariant restriction in (42). The superscript $(b)$ on {\it all}
the supervariables denotes that, in the super expansions of {\it these} supervariables, the coefficients of 
$\bar\theta$ are nothing but the off-shell nilpotent BRST symmetry transformations (13) for our theory.

We end this Appendix with the following remarks. First, the BRST and anti-BRST invariant restrictions 
on the (anti-)chiral supervariables [cf. Eqs. (40), (42), (46), (48)] are, precisely speaking, the
{\it quantum} gauge invariant restrictions which lead to provide the correct relationships among the 
{\it secondary} variables and {\it basic} as well as auxiliary variables of the (anti-)BRST invariant
theory. Second, the determination of the secondary variables [that lead to the derivation of the anti-BRST 
symmetry transformations (12) as the coefficients of $\theta$ in the expansion (50)] has been carried out following
exactly similar kind of procedure as we have done to derive the  relationship (43) from (42) in the case of determination
of the BRST symmetry transformations  (13) as the coefficient of $\bar\theta$ in the expansions (44). Finally, the ACSA to
BRST formalism is a simple but beautiful {\it symmetry-based} theoretical technique which is applicable to 
{\it all} kinds of physical systems of theoretical interest.\\

\begin{center}
{\bf Appendix D: On an Alternative Proof of the Absolute Anticommutativity and the Existence of CF-Type Restriction}\\
\end{center}
\vskip 0.5cm
The purpose of our present Appendix is to provide an {\it alternative} proof of the absolute anticommutativity of the conserved and 
nilpotent (anti-)BRST charges and the existence of the (anti-)BRST invariant [i.e. $s_{(a)b}\,(b + \bar b + 2\,\beta\,\bar\beta) = 0$] CF-type
restriction: $b + \bar b + 2\,\beta\,\bar\beta = 0$ in the {\it ordinary} space as well as in the {\it superspace} by exploiting the
theoretical potential and power of ACSA to BRST formalism. In this context, we note that the off-shell nilpotent $([Q_{(a)b}^{(2)}]^2 = 0)$
(anti-)BRST charges $Q_{(a)b}^{(2)}$ [cf. Eqs. (33), (30)] can be expressed as follows:
\[Q_{ab}^{(2)}  = s_{ab} \;\big[i\,(\bar c\,\dot c - \dot{\bar c}\,c) + 2\,i\,(\beta\,\bar c + \bar\beta\,c)\,\chi
 - 2\,e\,\bar\beta\beta + 2\,\gamma\,\psi_5 \big],\eqno (D.1)\]
\[Q_{b}^{(2)}  = s_{b}\; \big[i\,(\dot{\bar c}\,c - {\bar c}\,\dot c) - 2\,i\,(\bar\beta\,c + \beta\,\bar c)\,\chi
 + 2\,e\,\bar\beta\beta + 2\,\gamma\,\psi_5 \big],\eqno (D.2)\]
where the (anti-)BRST symmetry transformations have been quoted in Eqs. (13) and (12), respectovely. In other words, it is obvious that the BRST charge
$Q_b^{(2)}$ can be written as a BRST {\it exact} quantity and the anti-BRST charge $Q_{ab}^{(2)}$ can be re-expressed in the {\it exact}
form w.r.t. the anti-BRST symmetry transformation $s_{ab}$ [cf. Eq. (12)]. Thus, it is transparent now that we have the following 
explicit relationships, namely;
\[
~s_{ab} Q_{ab}^{(2)} = -\,i\,\{Q_{ab}^{(2)}, Q_{ab}^{(2)}\} = 0\;\, \Longleftrightarrow \;[Q_{ab}^{(2)}]^2 = 0\Longleftrightarrow s_{ab} ^2 = 0,\eqno (D.3)
\]
\[
s_{b} Q_{b}^{(2)} = -\,i\,\{Q_{b}^{(2)}, Q_{b}^{(2)}\} = 0\;\, \Longleftrightarrow \;[Q_{b}^{(2)}]^2 = 0\Longleftrightarrow s_{b} ^2 = 0,\eqno (D.4)
\]
where we have used the fundamental relationship between the continuous symmetry transformations $s_{(a)b}$ and their generators 
$Q_{(a)b}^{(2)}$. It is evident, from equations (D.3) and (D.4), that the off-shell nilpotency of the (anti-)BRST charges $Q_{(a)b}^{(2)}$
is deeply and intimately connected with the off-shell nilpotency of the (anti-)BRST symmetry transformations $s_{(a)b}$ in the {\it ordinary}
space. It is worthwhile to point out that these conclusions can {\it not} be drawn from our {\it earlier} proof of the off-shell nilpotency
[cf. Eqs.(38) and (36)].

In view of the mappings: $s_b \rightarrow \partial_{\bar\theta},\; s_{ab} \rightarrow \partial_\theta$, it is clear that the expressions
in (D.1) and (D.2) can be expressed in terms of the (anti-)chiral supervariables and the derivatives 
as well as the differentials w.r.t. the Grassmannian variables as:
\[
Q_{ab} ^{(2)} = \frac{\partial}{\partial\theta} \Big[i\,\Big(\bar F ^{(ab)} (\tau, \theta)\,\dot F ^{(ab)} (\tau, \theta) - 
\dot {\bar F} ^{(ab)} (\tau, \theta)\, F ^{(ab)} (\tau, \theta)\Big) + 2\,i\,\Big(\tilde\beta ^{(ab)} (\tau, \theta)\,\bar F ^{(ab)} (\tau, \theta)\]
\[ +  \tilde{\bar\beta}^{(ab)} (\tau, \theta)\,  F ^{(ab)} (\tau, \theta)\Big)\,\tilde \chi ^{(ab)} (\tau, \theta) -2\,E ^{(ab)} (\tau, \theta)\,\tilde{\bar\beta} ^{(ab)} (\tau, \theta)\,
\tilde\beta ^{(ab)} (\tau, \theta)\]
\[ + 2\,\Gamma ^{(ab)} (\tau, \theta)\,\Psi _5 ^{(ab)} (\tau, \theta) \Big]\]
\[~~~~~~~\equiv  \int d\theta \,\Big[i\,\Big(\bar F ^{(ab)} (\tau, \theta)\,\dot F ^{(ab)} (\tau, \theta) - 
\dot {\bar F} ^{(ab)} (\tau, \theta)\, F ^{(ab)} (\tau, \theta)\Big) + 2\,i\,\Big(\tilde\beta ^{(ab)} (\tau, \theta)\,\bar F ^{(ab)} (\tau, \theta)\]
\[ +  \tilde{\bar\beta}^{(ab)} (\tau, \theta)\,  F ^{(ab)} (\tau, \theta)\Big)\,\tilde \chi ^{(ab)} (\tau, \theta) -2\,E ^{(ab)} (\tau, \theta)\,\tilde{\bar\beta} ^{(ab)} (\tau, \theta)\,
\tilde\beta ^{(ab)} (\tau, \theta)\]
\[ + 2\,\Gamma ^{(ab)} (\tau, \theta)\,\Psi _5 ^{(ab)} (\tau, \theta) \Big], 
\eqno (D.5)\]
\[
Q_{b} ^{(2)} = \frac{\partial}{\partial\bar\theta} \Big[i\,\Big(\dot{\bar F} ^{(b)} (\tau, \bar\theta)\,F ^{(b)} (\tau, \bar\theta) - 
 {\bar F} ^{(b)} (\tau, \bar\theta)\, \dot F ^{(b)} (\tau, \bar\theta)\Big) - 2\,i\,\Big(\tilde{\bar\beta} ^{(b)} (\tau, \bar\theta)\,F ^{(b)} (\tau, \bar\theta)\]
\[ -  \tilde{\beta}^{(b)} (\tau, \bar\theta)\,\bar F ^{(b)} (\tau, \bar\theta)\Big)\,\tilde \chi ^{(b)} (\tau, \bar\theta) + 2\,E ^{(b)} (\tau, \bar\theta)\,\tilde{\bar\beta} ^{(b)} (\tau, \bar\theta)\,
\tilde\beta ^{(b)} (\tau, \bar\theta) \]
\[ + 2\,\Gamma ^{(b)} (\tau, \bar\theta)\,\Psi _5 ^{(b)} (\tau, \bar\theta) \Big]\]
\[\equiv \int d \bar\theta \,\Big[i\,\Big(\dot{\bar F} ^{(b)} (\tau, \bar\theta)\,F ^{(b)} (\tau, \bar\theta) - 
 {\bar F} ^{(b)} (\tau, \bar\theta)\, \dot F ^{(b)} (\tau, \bar\theta)\Big) - 2\,i\,\Big(\tilde{\bar\beta} ^{(b)} (\tau, \bar\theta)\,F ^{(b)} (\tau, \bar\theta)\]
\[ -  \tilde{\beta}^{(b)} (\tau, \bar\theta)\,\bar F ^{(b)} (\tau, \bar\theta)\Big)\,\tilde \chi ^{(b)} (\tau, \bar\theta) + 2\,E ^{(b)} (\tau, \bar\theta)\,\tilde{\bar\beta} ^{(b)} (\tau, \bar\theta)\,
\tilde\beta ^{(b)} (\tau, \bar\theta) \]
\[ + 2\,\Gamma ^{(b)} (\tau, \bar\theta)\,\Psi _5 ^{(b)} (\tau, \bar\theta) \Big].
\eqno (D.6)\]
It is crystal clear that, a close look at the above equations (D.5) and (D.6), 
leads to the following relationship within the framework of ACSA to BRST formalism:
\[
\partial_\theta \,Q_{ab}^{(2)} = 0\quad \Longleftrightarrow \quad \partial_{\theta}^{2} = 
0\quad \Longleftrightarrow \quad s_{ab}^{2} = 0,\eqno (D.7)\]
\[
\partial_{\bar\theta} \,Q_{b}^{(2)} = 0 \quad \Longleftrightarrow\quad  \partial_{\bar\theta}^{2}
= 0 \quad \Longleftrightarrow \quad s_{b}^{2} = 0.\eqno (D.8)\]
Thus, we have captured the off-shell nilpotency $[\big(Q_{(a)b}^{(2)}\big)^2 = 0]$ of the (anti-)BRST charges 
[cf. Eqs. (D.3), (D.4)] within the framework of ACSA and established that the off-shell nilpotency of the (anti-)BRST
symmetries and corresponding conserved charges is intimately connected with the nilpotency $(\partial_{\bar\theta}^{2}
= \partial_{\theta}^{2} = 0 )$ of the translational generators $(\partial_{\bar\theta},\,\partial_{\theta})$ along 
the Grassmannian $ (\bar\theta)\theta$-directions of the (1, 1)-dimensional (anti-)chiral super sub-manifolds
[of the general (1, 2)-dimensional supermanifold]. The relationships in (D.3) and (D.4) are sacrosanct
 and these are {\it expected} results within the framework of ACSA.

We now focus in the derivation of the CF-type restriction ($ b + \bar b + 2\,\beta\,\bar \beta = 0$) in the
{\it ordinary} space as well as in the {\it superspace } (by exploiting the basic techniques of ACSA). Toward
this goal in mind, first of {\it all}, we assume the {\it validity} of the CF-type restriction {\it on} our theory,
right from the beginning. As a consequence, we observe that the (anti-)BRST charges $Q_{(a)b}^{(2)}$ can be expressed in different forms as:
\[
Q_{ab}^{(3)}  = b\,\dot{\bar c} - \dot b\,\bar c - 2\,\bar\beta\,\dot\beta\,\bar c - 2\,\bar c\,\gamma\,\chi 
+ \bar\beta ^2 \,\dot c - 2\,i\,\bar\beta\,(m - e)\,\gamma  + 2\,b\,\bar\beta\,\chi + 2\,\bar\beta^2\,\beta\,\chi, \eqno (D.9)\]
\[~~Q_{b}^{(3)}  = \dot{\bar b}\,{c} - \bar b\,\dot c + 2\,\dot{\bar\beta}\,\beta\,c -  2\,c\,\gamma\,\chi 
- \beta ^2 \,\dot {\bar c} - 2\,i\,\beta\,(m - e)\,\gamma  - 2\,\bar b\,\beta\,\chi - 2\,\bar\beta\,\beta^ 2\,\chi. \eqno (D.10)\]
The above forms are very interesting because now we can express the BRST charge as an {\it exact} expression
w.r.t. the anti-BRST symmetry transformations (12) {\it and} the anti-BRST charge as an {\it exact} 
quantity w.r.t. the BRST symmetry transformation listed in (13). In other words, we have the following explicit expressions:
\[ Q_{ab}^{(3)} = s_b\, \bigl [i\,\dot{\bar c}\,\bar c - 2\,i\,\bar\beta\,\bar c\,\chi + e\,\bar\beta^2 - m\,\bar\beta^2 \bigr ], \eqno (D.11)
\]
\[Q_{b}^{(3)} = s_{ab}\,\bigl [i\,c\,\dot c + 2\,i\,\beta\, c\,\chi - e\,\beta^2 + m\,\beta^2 \bigr ]. \eqno (D.12)
\]
It is straightforward, at this juncture, to note that we have the following:
\[s_b Q_{ab}^{(3)}  = -\,i\,\{Q_{ab}^{(3)}, Q_{b}^{(3)}\} = 0 \quad\;\Longleftrightarrow \quad s_b ^2 = 0, \eqno (D.13)\]
\[~~~s_{ab} Q_{b}^{(3)}  = -\,i\,\{Q_{b}^{(3)}, Q_{ab}^{(3)}\} = 0 \quad\Longleftrightarrow \quad s_{ab} ^2 = 0. \eqno (D.14)\]
The above observations demonstrate that the absolute anticommutativity of the anti-BRST
charge {\it with} BRST charge is deeply connected with the off-shell nilpotency ($s_b^2 = 0$) of the 
BRST symmetry transformations (13). On the other hand, the absolute anticommutativity of
the conserved BRST charge {\it with} the anti-BRST charge is intimately connected with the off-shell nilpotency ($s_{ab}^2 = 0$)
of the anti-BRST symmetry
transformations (12). Thus, even in the {\it ordinary} space, a {\it close} look at (D.13) and (D.14) demonstrates that the proof of absolute
anticommutativity property {\it distinguishes} between the off-shell nilpotency of the BRST {\it and} anti-BRST symmetry transformations. This
exercise, in a {\it subtle} manner, {\it also} proves the existence of the anti-BRST invariant $[s_{(a)b}\,(b + \bar b + 2\,\beta\,\bar\beta)
= 0]$ CF-type restriction: $b + \bar b + 2\,\beta\,\bar\beta = 0$ on our theory of a 1D system of the reparameterization 
and (super)gauge invariant {\it massive} spinning relativistic particle. In this connection, it is worthwhile to point out that the {\it specific} forms
[cf. Eqs. (D.9), (D.10)] of the (anti-)BRST charges have been obtained from the {\it other} expressions for the (anti-)BRST charges [cf. Eqs. (33), (30)]  
{\it only} after the substitution of the CF-type restriction: $b + \bar b + 2\,\beta\,\bar\beta = 0$ of our theory.

We are now in the position to capture the {\it above} absolute anticommutativity property within the framework of ACSA. Keeping in our mind
the mappings: $s_b \leftrightarrow \partial_{\bar\theta},\; s_{ab} \leftrightarrow \partial_\theta$ (see, e.g., Refs. [10-12] for details), it is
evident that the observations in (D.11) and (D.12) can be translated into {\it superspace} (with the help of ACSA to BRST formalism) as:  
\[
Q_{ab}^{(3)} = \frac{\partial}{\partial\bar\theta}\,\Big[i\,\dot{\bar F}^{(b)} (\tau, \bar\theta)\,{\bar F}^{(b)} (\tau, \bar\theta) - 
2\,i\,\tilde{\bar\beta}^{(b)} (\tau, \bar\theta)\,\bar F^{(b)} (\tau, \bar\theta)\,\chi^{(b)} (\tau, \bar\theta)\]
\[ +
E ^{(b)} (\tau, \bar\theta) \,\tilde{\bar\beta}^{2(b)} (\tau, \bar\theta) - m\,\tilde{\bar\beta}^{2(b)} (\tau, \bar\theta) \Big]\]
\[~~~~~~~~\equiv  \int d\bar\theta\,\Big[i\,\dot{\bar F}^{(b)} (\tau, \bar\theta)\,{\bar F}^{(b)} (\tau, \bar\theta) - 
2\,i\,\tilde{\bar\beta}^{(b)} (\tau, \bar\theta)\,\bar F^{(b)} (\tau, \bar\theta)\,\chi^{(b)} (\tau, \bar\theta)\]
\[ +
E ^{(b)} (\tau, \bar\theta) \,\tilde{\bar\beta}^{2(b)} (\tau, \bar \theta) - m\,\tilde{\bar\beta}^{2(b)} (\tau, \bar\theta) \Big],
\eqno (D.15)\]
\[
Q_{b}^{(3)} = \frac{\partial}{\partial\theta}\,\Big[i\,{F}^{(ab)} (\tau, \theta)\,{\dot F}^{(ab)} (\tau, \theta) + 
2\,i\,\tilde{\beta}^{(ab)} (\tau, \theta)\,F^{(ab)} (\tau, \theta)\,\chi^{(b)} (\tau, \theta)\]
\[ -
E ^{(ab)} (\tau, \theta) \,\tilde{\beta}^{2(ab)} (\tau, \theta) + m\,\tilde{\beta}^{2(ab)} (\tau, \theta) \Big]\]
\[~~~~~~~~~\equiv \int d\theta\,\Big[i\,{F}^{(ab)} (\tau, \theta)\,{\dot F}^{(ab)} (\tau, \theta) + 
2\,i\,\tilde{\beta}^{(ab)} (\tau, \theta)\,F^{(ab)} (\tau, \theta)\,\chi^{(b)} (\tau, \theta)\]
\[ -
E ^{(ab)} (\tau, \theta) \,\tilde{\beta}^{2(ab)} (\tau, \theta) + m\,\tilde{\beta}^{2(ab)} (\tau, \theta) \Big],
\eqno (D.16)\]
where the superscripts $(b)$ and $(ab)$ on the {\it (anti-)chiral} supervariables have already been explained [cf. Eqs. (40), (44),
(46), (50)] earlier (see, Secs. 4 and 5). At this stage, we observe, in a straightforward fashion, that:
\[
\partial_{\bar\theta}\, Q_{ab}^{(3)} = 0 \quad \Longleftrightarrow \quad \partial_{\bar\theta}^{2} = 0 \quad \Longleftrightarrow \quad s_{b}^2 = 0,
\eqno (D.17)\]
\[
\partial_{\theta}\, Q_{b}^{(3)} = 0 \quad \Longleftrightarrow \quad \partial_{\theta}^{2} = 0 \quad \Longleftrightarrow \quad s_{ab}^2 = 0.
\eqno (D.18)
\]
The above equations are the reflections of our earlier observations in (D.13) and (D.14) which are nothing but 
the proof of absolute anticommutativity property of the conserved and nilpotent (anti-)BRST charges. In other
 words, we have been able to capture the absolute anticommutativity property within the purview of ACSA to BRST formalism.

We end this Appendix with the closing and key remarks that the proof of the absolute anticommutativity property 
of the (anti-)BRST charges within the framework of ASCA to BRST formalism {\it distinguishes} [cf. Eqs. (D.17), (D.18)]
 between the (1, 1)-dimensional {\it chiral} and {\it anti-chiral} super sub-manifolds of the {\it general} 
(1, 2)-dimensional supermanifold on which our 1D system of a massive spinning relativistic particle has been generalized.
This is due to the fact that the nilpotent ($\partial^2_{\bar\theta} = \partial^2_\theta = 0$) translational generators
($\partial_{\bar\theta}, \partial_\theta$) are defined separately and independently along the (anti-)chiral
Grassmannian directions of the (1, 1)-dimensional super sub-manifolds (of the (1, 2)-dimensional supermanifold).\\

\noindent
{\bf\large Data Availability}\\

\noindent
No data were used to support this study. \\

\noindent
{\bf\large Conflicts of Interest}\\

\noindent
The authors declare that there is no conflicts of interest.\\

\noindent
{\bf\large Acknowledgments}\\

\noindent
The present investigation has been carried out under the financial supports from BHU-fellowships
to A. Tripathi and A. K. Rao from Banaras Hindu University (BHU), Varanasi. B. Chauhan acknowledges 
the financial support from DST (Govt. of India) under {\it its} INSPIRE-fellowship
program. These authors express their deep sense of gratitude
to the above {\it local} and {\it national} funding agencies for {\it their} financial supports.\\


\begin{thebibliography}{99}
\bibitem{RPM1}     C. Becchi, A. Rouet, R. Stora, {\it Phys. Lett.} B {\bf 52}, 344 (1974) 
\bibitem{RPM2}     C. Becchi, A. Rouet, R. Stora, {\it Comm. Math. Phys.} {\bf 42}, 127 (1975) 
\bibitem{RPM3}     C. Becchi, A. Rouet, R. Stora, {\it Annals of  Physics}  {\bf 98}, 287 (1976) 
\bibitem{RPM4}     I. V. Tyutin, Lebedev Institute Preprint, Report Number: {\bf FIAN-39} (1975)\\ (unpublished),
                   arXiv: 0812.0580 [hep-th] 
\bibitem{RPM5}     L. Bonora, R. P. Malik, {\it Phys. Lett.} B {\bf 655}, 75 (2007)
\bibitem{RPM6}     L. Bonora, R. P. Malik, {\it J. Phys.} A: {\it Math. Theor.} {\bf 43}, 375403 (2010)
\bibitem{RPM7}     G. Curci, R. Ferrari, {\it Phys. Lett.} B {\bf 63}, 91 (1976)
\bibitem{RPM8}     J. Thierry-Mieg, {\it J. Math. Phys.} {\bf 21}, 2834 (1980)
\bibitem{RPM9}     M. Quiros, F. J. De Urries, J. Hoyos, M. L. Mazon, E. Rodrigues,\\
                   {\it J. Math. Phys.} {\bf 22}, 1767 (1981)              
\bibitem{RPM10}    L. Bonora, M. Tonin, {\it Phys. Lett.} B {\bf 98}, 48 (1981) 
\bibitem{RPM11}    L. Bonora,  P. Pasti,  M. Tonin, {\it Nuovo Cimento} A {\bf 64}, 307 (1981)
\bibitem{RPM12}    L. Bonora,  P. Pasti,  M. Tonin,  {\it Annals of Physics} {\bf 144}, 15 (1982)
\bibitem{RPM13}    R. Delbourgo, P. D. Jarvis, {\it J. Phys. A: Math. Gen.} {\bf 15}, 611 (1981)
\bibitem{RPM14}    L. Baulieu, J. Thierry-Mieg, {\it Nucl. Phys.} B {\bf 197}, 477 (1982)
\bibitem{RPM15}    L. Alvarez-Gaum´e, L. Baulieu, {\it Nucl. Phys.} B {\bf 212}, 255 (1983) 
\bibitem{RPM16}    R. P. Malik, {\it Eur. Phys. J.} C {\bf 60}, 457 (2009)
\bibitem{RPM17}    R. P. Malik,  {\it J. Phys. A: Math. Theor.} {\bf 39}, 10575 (2006)
\bibitem{RPM18}    R. P. Malik, {\it Eur. Phys. J.} C {\bf 51}, 169 (2007)
\bibitem{RPM19}    R. P. Malik, {\it J. Phys.  A: Math. Theor.} {\bf 37}, 5261 (2004) 
\bibitem{RPM20}    S. Kumar, B. Chauhan, R. P. Malik,  {\it Int. J. Mod. Phys.} A {\bf 33}, 1850133 (2018)
\bibitem{RPM21}    A. Shukla, N. Srinivas, R. P. Malik, {\it Annals of Physics} {\bf 394}, 98 (2018)
\bibitem{RPM22}    T. Bhanja, N. Srinivas,  R. P. Malik, {\it Int. J. Mod. Phys.} A {\bf 34}, 1950183 (2019) 
\bibitem{RPM23}    B. Chauhan, S. Kumar,  R. P. Malik, {\it Int. J. Mod. Phys.} A {\bf 34}, 1950131 (2019)
\bibitem{RPM24}    B. Chauhan, S. Kumar,  A. Tripathi,  R. P. Malik, \\{\it Adv. High Energy Phys.}  {\bf 2020}, 3495168 (2020)
\bibitem{RPM25}    A. Shukla, S. Krishna, R. P. Malik, {\it Eur. Phys. J.} C {\bf 72}, 2188 (2012)
\bibitem{RPM26}    B. Chauhan, A. Tripathi, A. K. Rao, R. P. Malik, arXiv: 1912.12909 [hep-th]
\bibitem{RPM27}    S. Krishna, D. Shukla, R. P. Malik, {\it Int. J. Mod. Phys.} A {\bf 28}, 1350108 (2018)
\bibitem{RPM28}    A. Shukla, T. Bhanja, R. P. Malik, {\it Eur. Phys. Lett.}, {\bf 101}, 51003 (2013)
\bibitem{RPM29}    R. P. Malik, {\it Int. J. Mod. Phys.} A {\bf 22}, 1053 (2007)
\bibitem{RPM30}    R. P. Malik, {\it Eur. Phys. J. C} {\bf 45}, 513 (2006)
\bibitem{RPM31}    R. P. Malik, {\it Int. J. Mod. Phys.} A {\bf 20}, 1767 (2005)
\bibitem{RPM32}    R. P. Malik, {\it J. Phys. A: Math. Theor.} {\bf 37}, 12077 (2004)
\bibitem{RPM33}    D. Nemschansky, C. Preitschopf, M. Weinstein, {\it Annals of Physics} {\bf 183}, 226 (1988)   
\bibitem{RPM34}    M. B. Green, J. H. Schwarz, E. Witten, {\it Superstring Theory}, Vols. {\bf 1} and {\bf 2}
                   \\(Cambridge University Press, Cambridge, 1987)
\bibitem{RPM25}    J. Polchinski, {\it String Theory} (Cambridge University Press, Cambridge, 1998)
\bibitem{RPM36}    P. A. M. Dirac, {\it Lectures on Quantum Mechanics}, 
                   Belfer Graduate School of Science\\
                   (Yeshiva University Press, New York, 1964)
\bibitem{RPM37}    K. Sundermeyer, {\it Constrained Dynamics: Lecture Notes in Physics}, 
                   Vol. {\bf 169},\\ (Springer-Verlag, Berlin, 1982)
\bibitem{Mir}      Mir Faisal, S. Upadhyay, {\it Phys. Lett.} B {\bf 736}, 288 (2014)
\bibitem{Mir2}     Mir Faisal, S. Upadhyay, B. P. Mandal, {\it Phys. Lett.} B {\bf 738}, 201 (2014)
\bibitem{Mir1}     S. Upadhyay, Mir Faisal, Prince A. Ganai, {\it Int. J. Mod. Phys.} A {\bf 30}, 1550185 (2015)
\bibitem{RPM38}    L. Bonora, R. P. Malik,  in preparation 
\bibitem{RPM39}    S. Weinberg, {\it The Quantum Theory of Fields}, Vol. {\bf II} \\(Cambridge University Press, Cambridge, 2005)




 
\end{thebibliography}
\end{document}